\documentclass[12pt,onecolumn]{IEEEtran}

\usepackage[OT1]{fontenc}
\usepackage{amsthm,amsmath,amssymb}
\usepackage[numbers]{natbib}
\usepackage{paralist}
\newcommand{\subparagraph}{}
\usepackage[compact]{titlesec}
\usepackage{setspace}

\usepackage{amsthm,amsmath,natbib}
\usepackage{mathrsfs}
\usepackage{algorithm}
\usepackage{algorithmic}
\usepackage{amssymb}
\usepackage{multirow}
\usepackage[english]{babel}
\usepackage{graphicx}
\usepackage{subfig}
\usepackage{url}

\newcommand{\bsb}{\boldsymbol}
\newcommand{\tr}{\mbox{tr}}

\newcommand{\tran}{{\mathsf{T}}}
\newcommand{\eqdef}{\triangleq}
\newcommand{\bsbA}{\bsb{A}}
\newcommand{\bsbB}{\bsb{B}}
\newcommand{\bsbC}{\bsb{C}}
\newcommand{\bsbG}{\bsb{G}}
\newcommand{\bsbx}{\bsb{x}}
\newcommand{\bsbX}{\bsb{X}}
\newcommand{\bsbY}{\bsb{Y}}
\newcommand{\bsbepsilon}{\bsb{\epsilon}}
\newcommand{\bsbOmega}{\bsb{\Omega}}
\newcommand{\bsbSigma}{\bsb{\Sigma}}

\newcommand{\bsbGamma}{\bsb{\Gamma}}

\begin{document}

%
%
%

\title{{Joint Association Graph Screening and Decomposition for Large-scale Linear Dynamical Systems}}
\author{Yiyuan She, Yuejia He, Shijie Li, and Dapeng Wu  \thanks{Yiyuan She is with Department of Statistics, Florida State University, Tallahassee, FL 32306. Yuejia He, Shijie Li and Dapeng Wu are with Department of Electrical and Computer Engineering,   University of Florida, Gainesville, FL 32611.
                  }}

\maketitle

\begin{abstract}
 This paper studies large-scale dynamical networks where the current state of the system is a linear transformation of the previous state, contaminated by a multivariate Gaussian noise. Examples include stock markets, human brains and gene regulatory networks. We introduce a transition matrix to describe the evolution, which can be translated to a directed Granger transition graph, and use the concentration matrix of the Gaussian noise to capture the second-order relations between nodes, which can be translated to an undirected conditional dependence graph. We propose  regularizing the two graphs jointly in topology identification and dynamics estimation. Based on the notion of joint association graph (JAG), we  develop a joint graphical screening and estimation (JGSE) framework for efficient network learning in big data. In particular, our method can pre-determine and remove unnecessary edges based on the joint graphical structure, referred to as JAG screening, and can decompose a large network into smaller subnetworks in a robust manner, referred to as JAG decomposition.  JAG screening and decomposition can reduce the problem size and search space for fine estimation at a later stage. Experiments on both synthetic data and real-world applications show the effectiveness of the proposed framework in large-scale network topology identification and dynamics estimation.
\end{abstract}

\begin{keywords}
Large-scale linear dynamical systems, graph learning, shrinkage estimation, variable selection.
\end{keywords}

\section{Introduction}
Topology learning and parameter estimation of dynamical networks have  become  popular research topics recently because such studies can reveal the underling mechanisms of many real-world complex systems. For example, a stock market which consists of a large number of stocks interacting with each other and evolving over time can be characterized as a dynamical network. Here, a node stands for the price of a stock and an edge or link resembles stock interaction.
Let $\boldsymbol{x}$ be a $p$-dimensional random vector with each
component being a time series associated with a node. We are interested in inferring the topology and  dynamics of a linear dynamical network
$\bsb{x}_t = \bsbA\bsb{x}_{t-1} + \bsbepsilon_t$, with $\bsbepsilon_t$ as the system disturbance.
Such a model has been proposed and studied in many areas such as econometrics, finance and bioinformatics~\cite{Sims1980, gourieroux2002financial, Fujita07, Bullmore09}. The transition matrix $\bsbA$ determines how the current state of the network evolves from the previous state.  It can be translated to a \textit{directed} Granger transition graph (\textbf{GTG}) that shows the Granger causal
connections  between the nodes~\cite{Granger69,s2,sigmoid}. The modern challenge is that
the number of unknowns\ in $\bsbA\in \mathbb{R}^{p\times p}$ is usually much larger
 than the number of available observations $\bsbx_1,\cdots,\bsbx_n$, i.e., $p^2\gg n$, and consequently most conventional  methods  fail in  estimation or identification.
From a statistical perspective, shrinkage estimation  \cite{Stein1956} must be applied, and  sparsity-promoting regularizations  are preferred because they can produce interpretable networks~\cite{bolstad2011causal, Pedro05}. Indeed,  in many applications, there only exist a few significant nodes that directly influence a given node. Sparse graph learning also complies with the principle of Occam's razor from a philosophical perspective.
Nevertheless, existing methods usually assume that the components of $\bsbepsilon_t$
are i.i.d., i.e., the covariance matrix of $\bsbepsilon_t$, or $Cov(\bsbx_t|\bsbx_{t-1})$, is proportional to the identity matrix. This totally ignores the \textit{second-order} statistical structure of the network. Most real-world networks violate this assumption because even conditioning on  past observations,  node correlations  widely exist.

Assuming, ideally, the true $\bsbA$ is known, the dependence structure of a network can be captured by the sparse Gaussian graph learning, which has attracted a lot of research attention lately (cf. \cite{banerjee2008model,Friedman08_graphLasso,Bickel2008,meinshausen2006high} among many others). Under $\bsbepsilon_t\sim N(\bsb{0}, \bsbSigma)$, the $(i,j)$th entry of the concentration matrix $\bsbOmega\eqdef \bsbSigma^{-1}$ gives the conditional dependence  between node $i$ and node $j$ given all the other nodes. This can be translated to an \textit{undirected} conditional dependence graph (\textbf{CDG}), in which case, again, sparsity on $\bsbOmega$ is desirable. Unfortunately, Gaussian graph learning  is not directly applicable to our  dynamical model, because as discussed above, the task of estimating $\bsbA$ is no less challenging as the task of estimating $\bsbOmega$. Note that  substituting the sample mean for the true mean is inappropriate when $\bsbA$ is a large matrix, which is a well known example of \textit{Stein's phenomenon}~\cite{Stein1956}.

To obtain a comprehensive picture of the dynamical network,
it is necessary to estimate both $\bsbA$ and $\bsbOmega$ based on their joint likelihood. There are few studies in the literature that consider the joint sparse estimation of the two matrices \cite{Rothman2010_SparseBsparseTheta, lee2012simultaneous}. In our experience, the existing methods are slow and can not handle big  network data. For example, the MRCE algorithm \cite{Rothman2010_SparseBsparseTheta} is already infeasible for  $p>120$ on an ordinary PC. Note that the number of unknown variables, $p^2+p(p+1)/2$, can be very large, thereby making it difficult to reliably identify the sparse network topology and accurately estimate the system parameters.

\begin{figure*}
  \centering
  \subfloat[\footnotesize sGTG]{\label{fig:betaConnect}\includegraphics[width=0.4\textwidth]{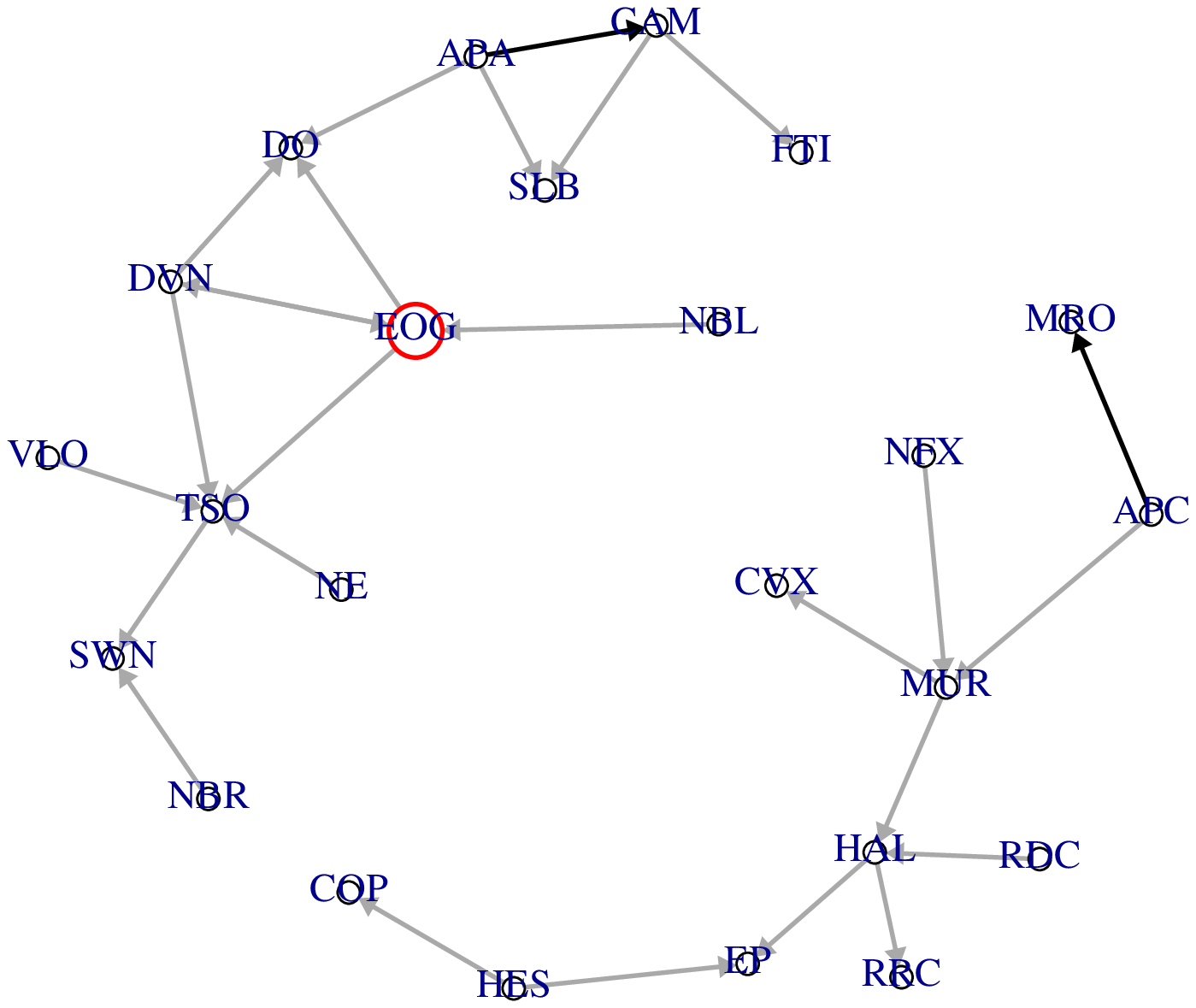}}
  \qquad
  \subfloat[\footnotesize sCDG]{\label{fig:omegaConnect}\includegraphics[width=0.4\textwidth]{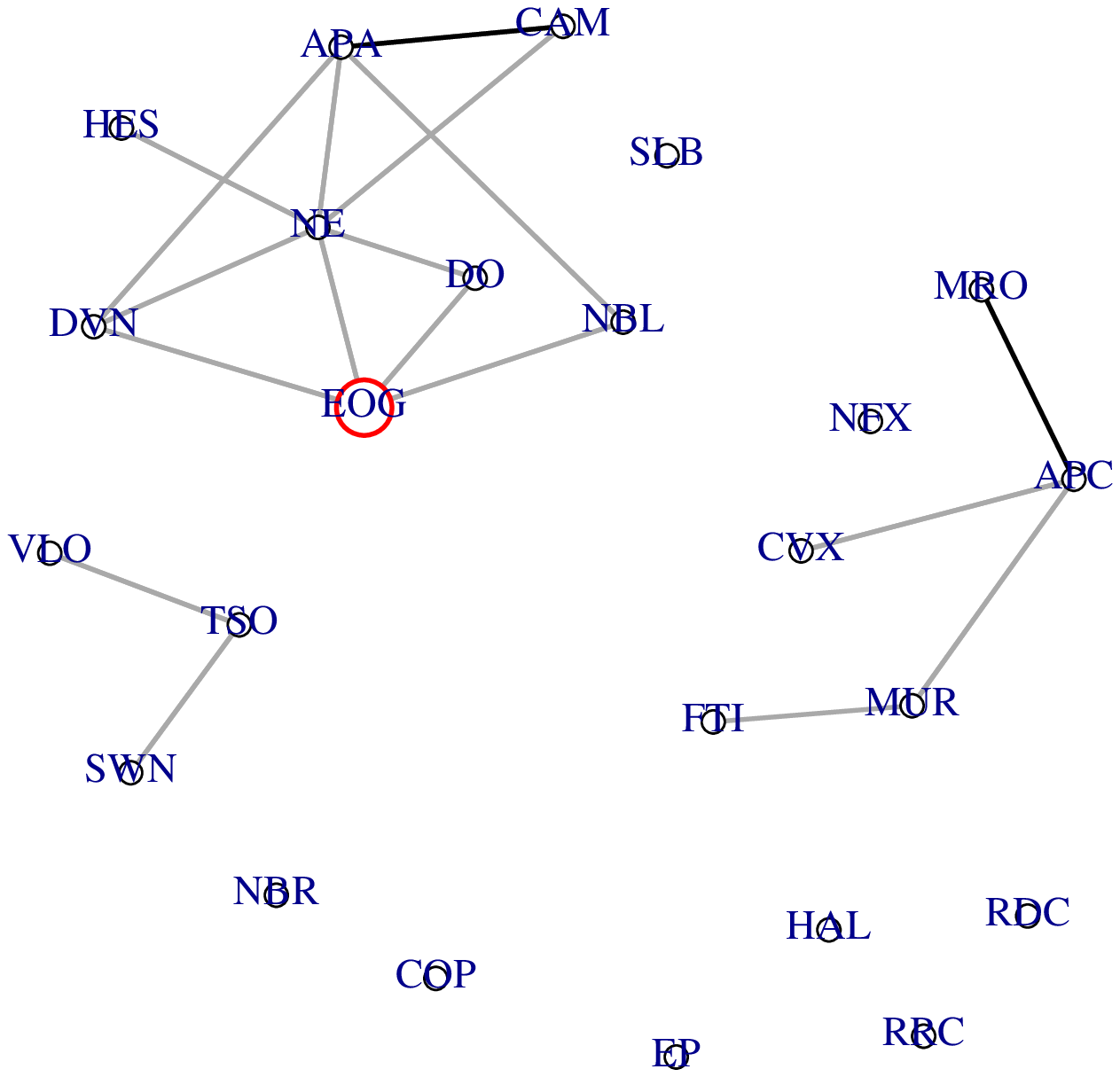}}
  \caption{\footnotesize The sGTG and sCDG for the $S\&P$ 500 indices from the ``Energy'' category. Common isolated nodes have been removed. Two nodes are connected with a line (\textit{directed} in GTG,  \textit{undirected} in CDG) if and only if their connection weight is nonzero. } \label{fig:ex_seperate_graphs}
\end{figure*}

As a real example, we use the Energy category  of the S\&P 500 stock data to illustrate our motivation.  Figure~\ref{fig:ex_seperate_graphs} shows the graphs obtained by sole GTG learning (sGTG for short) which ignores the second-order node correlations, and by  sole CDG learning (sCDG for short) which ignores the first-order Granger causalities. Common isolated nodes have been removed.  Some edges exist in both graphs, which suggests that  the \textbf{joint regularization} of $(\bsbA$, $\bsbOmega)$ might be helpful in detecting the joint graphical structure.
 In fact, statistically speaking, even when  similarities between the two graphs are not  clear or even do not exist, joint regularization can improve the overall estimation accuracy in  high dimensions, see, e.g., \cite{Stein1956,james1961}.
Another interesting observation from Figure~\ref{fig:ex_seperate_graphs} is that the network can be  decomposed into smaller subnetworks including isolated indices. Similar decomposability has also been noticed in brain connectivity networks \cite{Fink2009} and  U.S. macroeconomics \cite{stock2012generalized}. If such a network decomposition could be detected in an early stage, complex learning algorithms, such as MRCE and Gaussian graph learning, would apply in a much more efficient way (in a possibly parallel manner).
 Of course, the decomposition based on  sGTG or  sCDG alone may not be   trustworthy.
When $p$ is large and  both GTG and CDG are unknown, the graph screening/decomposition based on $\bsb{A}$ and $\bsbOmega$, jointly, is  much more reasonable.

This work proposes  jointly regularizing the directed transition graph and the undirected dependence graph for  topology identification and dynamics estimation.
We will introduce the notion of joint association graph (JAG) and propose  JAG screening and decomposition to facilitate large-scale network learning. The JAG screening identifies and removes unnecessary links.  The  JAG structure can also be used for network decomposition,   so that GTG or CDG can be estimated for each subnetwork separately.
With  search space and problem size substantially reduced, computational and statistical performance can be enhanced. 
Similar ideas have  proved to be very successful in Gaussian graph learning, such as  the block diagonal screening rule (BDSR)    \cite{Witten2011, Rahul2012}. Our approach is  based on JAG instead of CDG alone. Moreover, we will develop a robust  JAG decomposition that does not   incur  excessive estimation bias as BDSR does \cite{Zhao2012_huge}. Our approach does not   mask authentic  edges to guarantee decomposability.
To the best of our knowledge,  no work of \textit{joint} graph screening or decomposition  is available in the literature.

The remainder of this paper is organized as follows.
Section~\ref{sec:problem} describes the
joint graphical model and proposes a learning framework called \textit{joint graphical screening and estimation} (JGSE). Section~\ref{sec:stage1} develops an algorithm of  {graphical iterative screening via thresholding} to be  used for JAG screening and robust decomposition. Section~\ref{sec:stage2} gives
a  fine learning of graphs (FLOG) algorithm that estimates $\bsbA$ and $\bsbOmega$ after screening. In Section~\ref{sec:experiment}, synthetic-data experiments are conducted to show the performance
of JGSE. In
Section~\ref{sec:application}, we apply JGSE to
real  S\&P 500  and  NASDAQ-100 stock data for network learning. 

\section{The joint graphical model}
\label{sec:problem}
Suppose there exist $p$ nodes in a dynamical network and let $\bsbx$ be a $p$-dimensional random vector with each
component  associated with a node. To describe  the node behaviors at each time point, we define  a linear  dynamical network model
\begin{equation}
\bsbx_{t}=\bsbA\bsbx_{t-1}+\bsbepsilon_{t},  \quad \bsbepsilon_t\sim
\mathcal{N}(\bsb{0},\bsbSigma).
\label{eq:model}
\end{equation}
The current state of the system is determined by  two components: The first component is a linear transformation of the previous state;  the second component $\bsbepsilon_t$  follows a multivariate Gaussian distribution and characterizes node correlations conditioned on  $\bsbx_{t-1}$.  The transition matrix $\bsbA$  can be translated to a directed \emph{Granger transition graph} (GTG): $a_{ij}\neq 0$ indicates that node $j$ Granger-causes node $i$~\cite{Granger69}.  The
concentration matrix $\bsb{\Omega}\eqdef \bsb{\Sigma}^{-1}$
can be translated to an undirected \textit{conditional dependence
graph} (CDG): $\omega_{ij}=\omega_{ji}=0$ indicates that node $i$ and node $j$ are conditionally independent given  the other nodes \cite{banerjee2008model,Friedman08_graphLasso} and $\bsbx_{t-1}$. Given $n+1$ snapshots of the system, $\bsbx_1,\cdots,\bsbx_{n+1}$, we would like to recover  the first-order statistic $\bsbA$ and  the second-order statistic $\bsbOmega$ as well as find out  their sparsity patterns (or  topological structures).
We are particularly interested in dynamical networks with both GTG and  CDG being \textbf{sparse} or approximately sparse for the following reasons.  First, many real-world  dynamical networks are indeed sparse. For example, in  regulatory networks, a gene is only regulated by several other genes~\cite{Fujita07}. Second, when the number of observations is small compared with the number of unknown variables, the sparsity assumption reduces the number of model parameters so that estimating  the system becomes possible. Third, from a philosophical point of view, a sparse model is consistent with the principle of Occam's razor and is easier to interpret in practice.

As  pointed out by a referee, sCDG which estimates $Cov(\bsbx_t)$ typically yields a less sparse graph than $\bsbSigma^{-1}$, because the transition matrix $\bsbA$, together with the autoregressive mechanism, brings in further node dependence  (see, e.g., \cite{LutkBook} for more details).
\subsection{Joint regularization in network learning}
\label{sec:sparseLearningofAandOmega}
Using the Markov property and
chain rule, we can write the joint likelihood of
$\bsbA$ and $\bsbOmega$ (conditioned on $\bsbx_1$) as
$
l(\bsbA,\bsbOmega)  = f(\bsbx_{n+1},\cdots,\bsbx_2|\bsbx_{1},\bsbA, \bsbOmega)  =\prod_{t=1}^n f(\bsbx_{t+1}|\bsbx_{t},\bsbA, \bsbOmega)$.
So the joint  ML estimation solves
$
\min_{\bsbA,\bsbOmega\succ 0} \frac{1}{2}\sum_{t=1}^n
(\bsbx_{t+1}-\bsbA\bsbx_{t})^\tran \bsbOmega
(\bsbx_{t+1}-\bsbA\bsbx_{t})-\frac{n}{2}\log
|\bsbOmega|. \label{eq:ML}
$
Let $\bsbY=[\bsbx_2,\cdots,\bsbx_{n+1}]^\tran,
\bsbX=[\bsbx_1,\cdots,\bsbx_{n}]^\tran$ and
$\bsbB=\bsbA^\tran$. We write the  ML problem  in matrix form
\begin{align}
\min_{\substack{
   \bsbB,\bsbOmega\in \mathbb{R}^{p\times p}\\
   \bsbOmega\succ 0
  }}   L(\bsbB,\bsbOmega)  =\frac{1}{2}\tr\{(\bsb{Y}-\bsb{X}\bsb{B})
\bsb{\Omega}(\bsb{Y}-\bsb{X}\bsb{B})^{\tran}\} -\frac{n}{2}\log
|\bsb{\Omega}|.
\label{eq:MLmatrix}
\end{align}
Here $\bsbOmega\succ 0$ means that $\bsbOmega$ is positive definite (which implies that $\bsbOmega$ is symmetric).
From now on, we use $\bsbB$, in place of $\bsbA$, to
represent the GTG.

To enforce sparsity, a straightforward idea is to add penalties, $P_B(\bsbB;\lambda_{B})$ and $P_{\Omega}(\bsb{\Omega};\lambda_{\Omega})$,  to the loss in \eqref{eq:MLmatrix}. $P_B$ and $P_{\Omega}$ can be of the  $\ell_1$ type \cite{Tibshirani96,Yuan07,Rothman2010_SparseBsparseTheta}.  In this paper, we propose  \textbf{jointly regularizing} $\bsbB$ and $\bsbOmega$  via penalty $P_C(\bsbC(\bsbB, \bsb{\Omega}); \lambda_{C})$, where $\bsbC$ is constructed based on $\bsbB$ and $\bsbOmega$. This  leads to the following   optimization problem
\begin{equation}
\min_{\bsbB,\bsb{\Omega}\succ 0}
L(\bsbB,\bsb{\Omega})+P_C(\bsbC(\bsbB, \bsb{\Omega});
\lambda_{C})+P_B(\bsbB;\lambda_{B})+P_{\Omega}(\bsb{\Omega};\lambda_{\Omega}).
\label{eq:all_penalty}
\end{equation}
The design of $P_C(\bsbC(\bsbB, \bsb{\Omega});\lambda_C)$ is to capture the  joint structure of  GTG and  CDG. Of course, joint regularization can reinforce the  detection of common edges if they exist.  {But why does one care about  the joint graphical structure in computation and statistics?} Some motivations are given below.

1) First, due to the sparsity assumption on $\bsbA$ and $\bsbOmega$, the union of the two  graphs is still sparse. That is, many  nodes have no direct influences. Hence  one can  perform  graph screening in an earlier stage for dimension reduction, to facilitate  fine GTG and CDG learnings afterwards. A good screening process should  take both graphs into account in  removing unnecessary hypothetical edges.


2) Many very large dynamical networks
demonstrate smaller-scale subnetwork structures or clusters. For example, a human brain connectivity network revealed by EEG data can be divided into several functionality regions~\cite{Fink2009}. Also, in the U.S. macroeconomic network,  economic indices can be divided to different categories~\cite{stock2012generalized}. It is desirable to decompose a large-scale network into small subnetworks, if possible, for both computational and statistical concerns ~\cite{Witten2011, Rahul2012}. In the dynamical network setting, such a decomposition must be based on both GTG and CDG.

3) Finally, the joint regularization  helps improve the overall identification and estimation accuracy based on some classical statistics literature  \cite{Stein1956,james1961,Efron1973}.



\subsection{Joint association graph}
Model \eqref{eq:model} shows the network evolves through both  first-order and  second-order statistical relations between the nodes.  To capture the joint structure, we introduce the notion of \textit{joint association graph} (JAG), an {undirected} graph where any two nodes are connected if they are connected in either GTG or CDG.
Define the ``association strength'' between node $i$ and node $j$  as
\begin{equation}
c_{ij}= c_{ji} = \sqrt{b_{ij}^2+b_{ji}^2+
 2\phi^2 \omega_{ij}^2}, \label{eq:defineC}
\end{equation}
where $\phi$ is a weight parameter (say, $\phi=1$); the matrix $\bsbC=[c_{ij}]\in\mathbb{R}^{p\times p}$  represents the JAG.

\begin{figure*}
  \centering
  \subfloat[GTG]{\label{fig:causalEX}\includegraphics[width=0.25\textwidth]{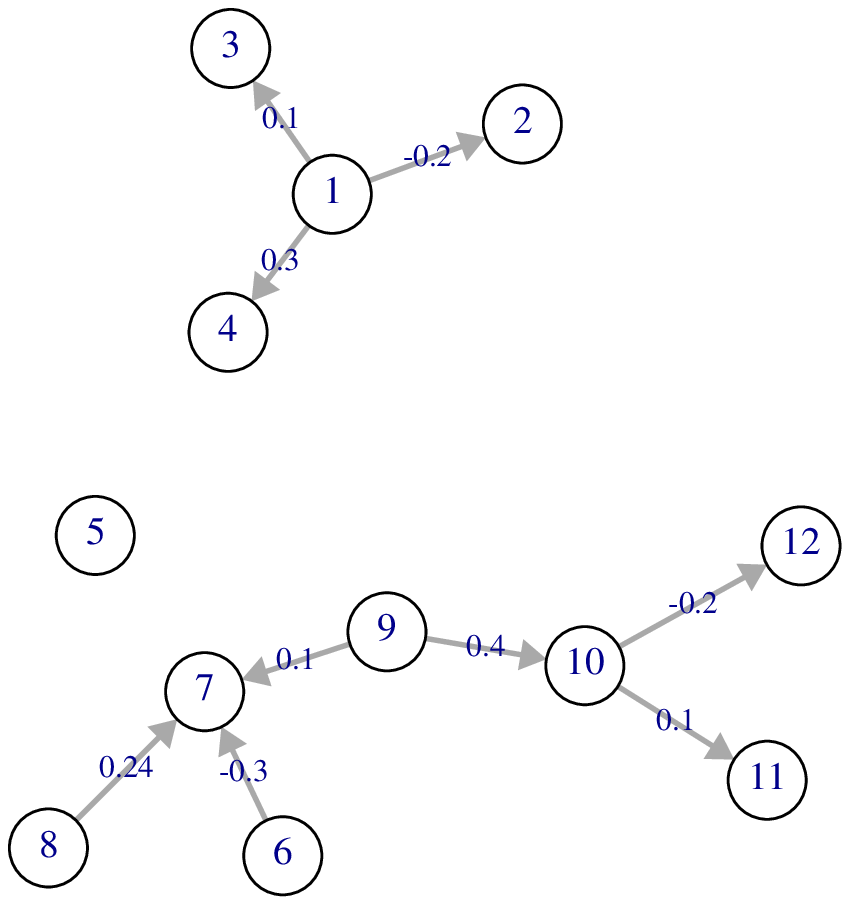}}
  \subfloat[CDG]{\label{fig:dependenceEX}\includegraphics[width=0.25\textwidth]{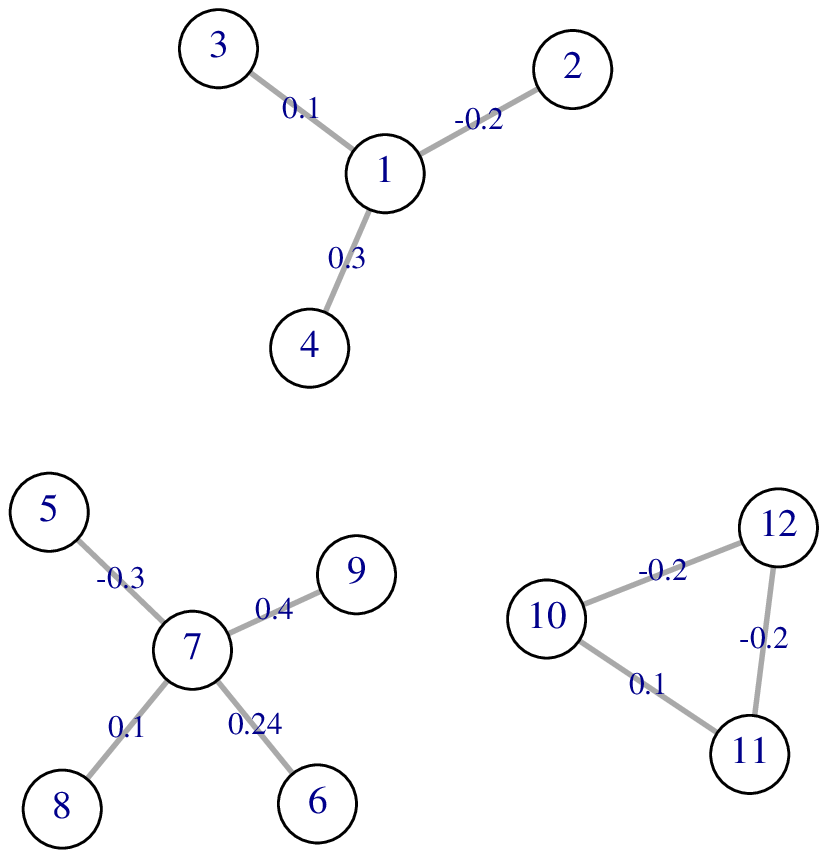}}
\subfloat[JAG]{\label{fig:jointEX}\includegraphics[width=0.25\textwidth]{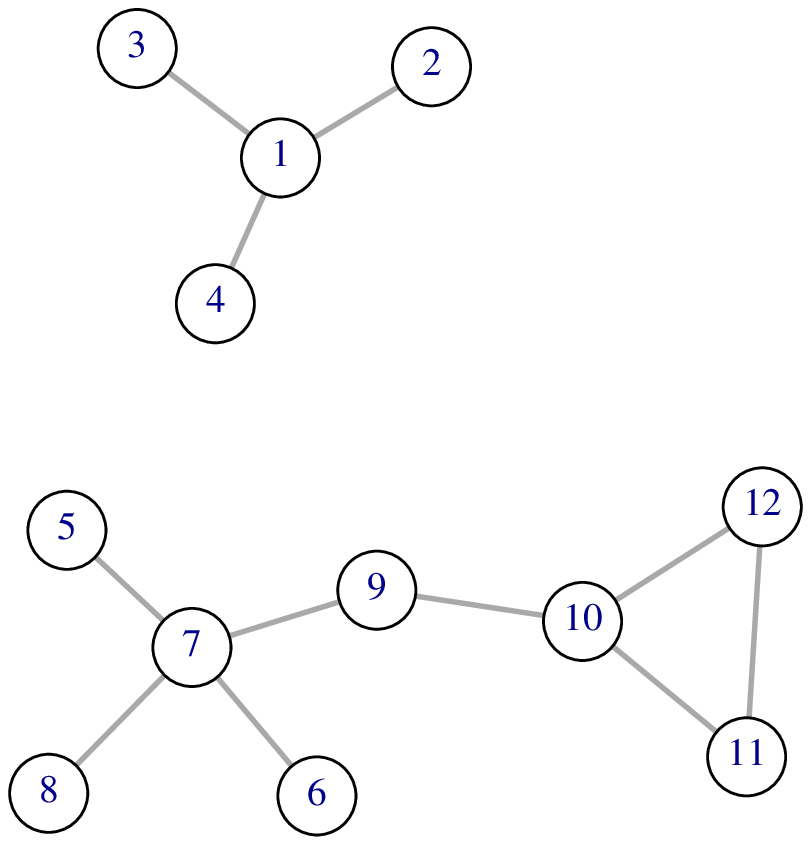}}
  \caption{\footnotesize{Demonstration of the joint graphical model.}} \label{fig:EXofJAG}
\end{figure*}

To give an illustration of JAG, we show a toy example in Figure~\ref{fig:EXofJAG}, where the JAG in Figure~\ref{fig:jointEX} is obtained from \eqref{eq:defineC}. The GTG and CDG  share many common edges. Furthermore, they both exhibit subnetwork structures. In fact,  in both graphs, nodes 1-4 form a cluster. On the other hand, the two graphs  differ from each other in some significant ways. For example, node 9 and node 10 are disconnected in GTG,  but not so in CDG. JAG, by  integrating the connections in GTG and CDG,  provides a comprehensive picture of  the network topology.

In reality, both the
GTG and CDG are unknown and to be
estimated. If one had the JAG learned
beforehand,  its structure could be used to perform graph screening and help improve the
estimation of $\bsb{B}$ and $\bsb{\Omega}$. For example, in Figure~\ref{fig:jointEX}, node 4 and node 5 are disconnected, so  setting $b_{45}=b_{54}=\omega_{45}=\omega_{54}=0$ beforehand facilitates network estimation and identification. Particularly, if the
JAG, after permutation, exhibits a
block-diagonal structure---$\mbox{diag}\{\bsb{C}_{11}, \cdots, \bsb{C}_{dd}\}$,
then both $\bsb{B}$ and $\bsb{\Omega}$ must have the same
block-diagonal structure, $\mbox{diag}\{\bsb{B}_{11}, \cdots, \bsb{B}_{dd}\}$ and $\mbox{diag}\{\bsb{\Omega}_{11}, \cdots, \bsb{\Omega}_{dd}\}$, respectively.
%
 It is not difficult to show that such a network can be decomposed
 into $d$ independent subnetworks with its dynamics properties completely intact. For example,
the network shown in Figure~\ref{fig:EXofJAG} can be decomposed into
two mutually disconnected subnetworks according to its JAG. We can estimate and  infer
GTG and CDG for each  subnetwork separately. Explicitly estimating  the JAG based on \eqref{eq:defineC} also facilitates  computation and algorithm design, as will be shown in Section~\ref{sec:JAG}.

\subsection{The JGSE learning }
\label{sec:two-stage}
Directly tackling  the jointly regularized problem  \eqref{eq:all_penalty}  is  extremely challenging.
(Even without $P_C$, the existing algorithms are inefficient or even infeasible for moderate-scale problems.) The key of the paper is to detect and utilize the joint structure of $\bsbB$ and $\bsbOmega$  for  computational and statistical performance boost.  We propose a \textbf{Joint Graphical Screening and Estimation} (JGSE) learning framework which consists of two stages: \textit{1) JAG screening and decomposition;  2) fine estimation}.

In Stage 1, we identify the  structure of JAG by solving the joint regularization problem
\begin{equation}
\min_{\bsbB,\bsb{\Omega}\succ 0} f_C(\bsbB,\bsb{\Omega}; \lambda_C)= L(\bsbB,\bsb{\Omega}) +
 P_C(\bsbB,\bsb{\Omega}; \lambda_C),
 \label{eq:stage1}
\end{equation}
where the penalty (or constraint) takes the form
\begin{equation}
P_C(\bsbB,
 \bsb{\Omega};\lambda_C)= \sum_{\substack{1\leq i <  j\leq p}} P( \sqrt{b_{ij}^2+b_{ji}^2+
 2\phi^2
\omega_{ij}^2 }; \lambda_C). \label{eq:groupPenalty}
\end{equation}
That is, we place $(b_{ij}, b_{ji}, \omega_{ij})$ into the same group, and use a sparsity-inducing penalty at the group level. The group sparsity pursuit ensures that as long as any type of connection between node $i$ and node $j$ exists, the    group will be kept, and so will the corresponding edge in  JAG. The grouping of variables can be arbitrary. Our algorithms apply provided that the groups do not overlap.
For
example, if we know a priori that several nodes form a cluster, we
can put the corresponding elements of  $\bsbB$ and
$\bsb{\Omega}$ into a group. The
form of \eqref{eq:groupPenalty} serves for the
general case where no particular prior information  is given.

Stage 2  estimates $(\bsbB, \bsb{\Omega})$ finely, given the pattern of  JAG:
\begin{equation}
\begin{gathered}
\min_{\bsbB,\bsb{\Omega}\succ 0}
L(\bsbB,\bsb{\Omega})+P_B(\bsbB;\lambda_B)+P_{\Omega}(\bsb{\Omega};\lambda_{\Omega}),\\
\text{ s.t. } E_B \subset E_{\hat{C}}, E_{\Omega} \subset
E_{\hat{C}}
\end{gathered} \label{eq:stage2}
\end{equation}
where $E_{\hat C}$ denotes the set of  nonzero edges in $\hat C$ and $E_B, E_C$ are similarly defined.
The constraints maintain the sparsity structure of $\bsb{\hat{C}}$ learned from Stage 1. 
 In this fine estimation stage, the number
of variables to be estimated has been substantially reduced. Packages for sparse matrix operations can be used. When JAG decomposition is possible, popular graph learning algorithms can be directly applied to
subnetworks,  and parallelism can be employed for high performance computing.

Both  sGTG   and  sCDG   are special cases of \eqref{eq:all_penalty}, and  can  be learned by screening + fine estimation as well. 
Ignoring the second-order network structure and assuming $\bsbepsilon$ has i.i.d. components, i.e., $\bsbOmega = \bsb{I}/\sigma^2$, the joint graphical model degenerates to the sGTG model where a sparse transition matrix $\bsbB$ can be obtained by solving
$
\bsb{\hat{B}} = \arg\min_{\bsb{B}} \frac{1}{2}\|\bsbY-\bsbX\bsbB\|_2^2
 +
P_B(\bsb{B};\lambda_{B}).\label{eq:iGCG}
$
Assuming the data have been centered and $\bsbB=\bsb{0}$, the joint graphical model degenerates to the sCDG model where a sparse $\bsbOmega$ can be obtained by Gaussian graph learning
$
\boldsymbol{\hat{\Omega}} = \arg\min_{\bsb{\Omega}\succ 0}
\tr\{\boldsymbol{S}\bsb{\Omega}\}-\log|\bsb{\Omega}|+\frac{2}{n}P_\Omega(\bsbOmega; \lambda_\Omega),
\label{eq:sCDGG}
$
with $\bsb{S}=\bsbY^\tran \bsbY/n$.
Another instance is the multivariate
regression with covariance estimation (MRCE)~\cite{Rothman2010_SparseBsparseTheta}:
$
\min_{\bsb{B},\bsb{\Omega}\succ 0}  L(\bsbB,\bsbOmega)
 +
P_B(\bsb{B};\lambda_B)+P_{\Omega}( \bsb{\Omega};\lambda_{\Omega})$. MRCE estimates both $\bsbB$ and $\bsbOmega$ but imposes no joint regularization. In our experience   MRCE is only  feasible for small-scale network learning, which is a main motivation of our JAG screening.
In the following two sections, we present computational algorithms for the two-stage JGSE learning framework. 

\section{JAG screening and decomposition}
\label{sec:stage1}
The objective function \eqref{eq:stage1} is nonconvex and nonsmooth and there are  a large number of unknown variables.
 One possible way to minimize \eqref{eq:stage1} is to use  coordinate descent; 
the resulting algorithm design is however quite cumbersome---one must consider different cases depending on whether the variables appearing in \eqref{eq:groupPenalty} are zero or not. Our experiments show that such an algorithm is only feasible for
 $p< 120$. More efficient
algorithms are in great need.  We
propose a novel GIST algorithm based on the group $\Theta$-estimator with asynchronous Armijo-type line search.

\subsection{Group $\Theta$-estimator}
\label{sec:group_theta}
To solve \eqref{eq:stage1}, we start from thresholding rules rather than penalty functions, considering that different penalty forms  may result in the same estimator (and the same thresholding operator)~\cite{She09}. 

A thresholding rule $\Theta(\cdot;
\lambda)$ is required to be an odd
nondecreasing unbounded shrinkage function \cite{SheGLMTISP}. Examples include the soft-thresholding operator $\Theta_S(t;\lambda)=\mbox{sgn}(t)(|t|-\lambda)1_{|t|> \lambda}$ and the hard-thresholding
 $\Theta_H(t;\lambda)=t1_{|t|> \lambda}$. (Throughout the paper, the sign function is defined as $\mbox{sgn}(t) = 1$ if $t>0$, $-1$ if $t<0$, and 0 if $t=0$.) When $\bsb{t}$ is a  vector, the thresholding rule  is defined componentwise. The  \textbf{multivariate} version of  $\Theta$, denoted by
$\vec{\Theta}(\bsb{t}; \lambda)$,  is defined by 
\begin{equation}
\vec{\Theta}(\bsb{t}; \lambda) = \bsb{t}^{\circ} \Theta(\|\bsb{t}\|_2;
\lambda), \text{ where } \bsb{t}^{\circ} = \begin{cases}
\frac{\bsb{t}}{\|\bsb{t}\|_2} & \text{ if } \bsb{t}\neq \bsb{0} \\
\bsb{0} & \text{ otherwise}.
\end{cases}
\label{eq:vector_th_rule}
\end{equation}

Now we formulate a general framework for solving  a general group penalized problem
\begin{equation}
\label{eq:gp}
\min_{\bsb{\beta}} -l(\bsb{\beta})+\sum_{k=1}^K P_k(\|\bsb{\beta}_k\|_2;\lambda_k),
\end{equation}
where $l$ is the log-likelihood function, and  $P_k$ are penalty functions possibly nonconvex (and discrete) with corresponding thresholding rules as described in \eqref{eq:construction}.  \cite{SheGLMTISP} shows that given any thresholding operators $\Theta_1, \cdots \Theta_K$,  the  iterative multivariate thresholding procedure
$
\bsb{\beta}_k^{l+1}=\vec{\Theta}_k(\bsb{\beta}_k^l-\alpha \frac{\partial l(\bsb{\beta})}{\partial \bsb{\beta}_k}|_{\bsb{\beta}=\bsb{\beta}^l}; \lambda_k)
$ for $1\leq k \leq K$
is guaranteed to converge under a universal choice of $\alpha$; moreover,  the convergent solution (referred to as a  \textit{group $\Theta$-estimator}) is a local minimum point of \eqref{eq:gp}, provided that $P_k$ and $\Theta_k$ are coupled through the following equation:
\begin{equation}
P_k(t;\lambda_k)-P_k(0;\lambda_k) = \int_0^{|t|}( \mbox{sup}\{s: \Theta_k(s;\lambda_k)\leq u\}-u )du
+ q_k(t;\lambda_k)
\label{eq:construction}
\end{equation}
for some nonnegative $q_k(\cdot;\lambda_k)$  satisfying $q_k(\Theta_k(s;\lambda_k); \lambda_k)=0$, $\forall s$.
We emphasize that the conclusion holds for \textit{any} thresholding rules, and most practically used penalties (convex or nonconvex) are covered by \eqref{eq:construction}. Two important examples that will play an important role  in the work are given as follows. When all $\vec\Theta_k$ take the form of   group soft-thresholding $\vec{\Theta}_S(\cdot;\lambda)$, the  corresponding penalty in \eqref{eq:gp} becomes the group $\ell_1$  penalty $\lambda\sum_{k=1}^K\|\bsb{\beta}_k\|_2$. When  all $\vec\Theta_k$ are chosen to be  group hard-thresholding $\vec{\Theta}_H(\cdot;\lambda)$, \eqref{eq:construction} yields  infinitely many  penalties  even when  $P(0; \lambda)=0$, one of which is  the  exact group  $\ell_0$ penalty $\sum {\lambda^2}1_{\|\bsb{\beta}_k\|_2\neq 0}/2$ by setting $q(t;\lambda)= 0.5 (\lambda - |t|)^2 1_{0<|t|<\lambda}$.

We now use the group $\Theta$-estimator to deal with problem  \eqref{eq:stage1}. Divide the variables in
$\bsbB$ and $\bsb{\Omega}$ into $K=p(p+1)/2$ groups, where variables at entry $(i,j)$ and entry $(j,i)$ ($1\leq i \leq j \leq p$)
belong to the $k$th group with $k=(i, j)$. Let $\bsb{\Gamma} = [\bsbB, \phi\bsbOmega]\in \mathbb{R}^{p\times 2p}$. It is not difficult to compute  the gradients of $L(\bsbB,\bsbOmega)$ with respect to $\bsbB$ and $\bsbOmega$   (details omitted)
\begin{equation}
\begin{aligned}
 \nabla_{\bsbB} L =
(\bsb{X}^\tran\bsb{X}\bsbB-\bsb{X}^\tran\bsb{Y})\bsb{\Omega}\eqdef\bsb{G}_B, \\
\nabla_{\bsbOmega} L =
\frac{1}{2}(\bsb{Y}-\bsb{X}\bsbB)^\tran(\bsb{Y}-\bsb{X}\bsbB) -
\frac{n}{2} \bsb{\Omega}^{-1}  \eqdef \bsb{G}_{\Omega}.
\end{aligned}
\label{eq:partialDerivatives}
\end{equation}
Thus the
gradient of $\bar L(\bsb{\Gamma})\eqdef L(\bsbB, \bsbOmega)$ with respect to $\bsb{\Gamma}$ is
\begin{equation}
 \nabla_{\bsb{\Gamma}} \bar L = [\bsb{G}_B, \phi^{-1}\bsb{G}_{\Omega}] \eqdef \bsb{G}.
\label{eq:derivative}
\end{equation}

Given $1\leq i\leq j\leq p$, let   $\bsb{\Gamma}_k=[\gamma_{ij}, \gamma_{ji}, \gamma_{i(j+p)},
\gamma_{j(i+p)}]^\tran$ or $[b_{ij}, b_{ji}, \phi\omega_{ij},
\phi\omega_{ji}]^\tran$, consisting  of all elements in $\bsb{\Gamma}$ that
belong to the $k$th group, and similarly, let  $\bsb{G}_k=[g_{ij}, g_{ji},
g_{i(j+p)}, g_{j(i+p)}]^\tran$. We extend the multivariate thresholding to such matrices. Given any thresholding $\Theta$, define its  multivariate  thresholding  $\vec{\Theta}(\bsbGamma; \lambda)$  as a new matrix $\tilde{\bsbGamma}$ satisfying  $\tilde{\bsbGamma}_k=\vec{\Theta}(\bsbGamma_k;\lambda)$, $\forall k$, with $\vec{\Theta}$ given by \eqref{eq:vector_th_rule}.
Then, the iterative algorithm to get a  group $\Theta$-estimator
of \eqref{eq:stage1} becomes
\begin{equation}
\bsb{\Gamma}^{l+1} \leftarrow \vec{\Theta} (\bsb{\Gamma}^l -
\alpha^l \bsb{G}^l; \lambda_C),
\label{eq:group_th}
\end{equation}
with $(P, \Theta)$ coupled through \eqref{eq:construction}.

\subsection{JAG screening}
\label{sec:JAG}
Equation \eqref{eq:group_th} can deliver a local minimum to problem \eqref{eq:stage1} for any penalty function constructed from a thresholding rule via  \eqref{eq:construction}.
This covers   $\ell_0, \ell_1$,  SCAD~\citep{Fan01_SCAD}, $\ell_p$ ($0<p<1$), and many other penalties \cite{SheGLMTISP}. The problem now boils down to choosing a proper penalty form for JAG screening. Another issue that cannot be ignored is parameter tuning, which is a nontrivial task especially for nonconvex  penalties.

Among all  sparsity-promoting penalties, it is of no doubt that the group $\ell_0$ penalty is  ideal in enforcing sparsity. However, its parameter tuning is not easy, and most tuning approaches, e.g.,  cross validation, become prohibitive in large network applications. Rather than using the group $\ell_0$ penalty, we propose using a group $\ell_0$ constraint for JAG screening
\begin{equation}
\label{eq:l0_constraint}
\sum_{1\leq i< j\leq p} 1_{(b_{ij},b_{ji},\omega_{ij})\neq \bsb{0}}\leq m.
\end{equation} This particular $\ell_0$ form enables one to directly control the cardinality\footnote{The cardinality of a network  refers to the number of nonzero links in $\bsbC$ in this paper.} of the network. (Note that the constraint excludes  the diagonal entries of $\bsbB$ and $\bsbOmega$.) The upper bound $m$ can be loose for the JAG screening step. This group $\ell_0$ constrained problem can be solved using the technique  in Section~\ref{sec:group_theta}, resulting in   a \textit{quantile} version of  \eqref{eq:group_th}:
\begin{equation}
\bsb{\Gamma}^{l+1} \leftarrow \vec{\Theta}^\# (\bsb{\Gamma}^l -
\alpha^l \bsb{G}^{l}; m).
\label{eq:quantile_th}
\end{equation}
Here, the multivariate quantile thresholding operator $\vec{\Theta}^\# (\cdot; m)$  \cite{SheSpec} for any $\bsbGamma\in \mathbb{R}^{p\times 2p}$ is defined to be a new matrix $\tilde{\bsbGamma}$ with $\tilde{\bsbGamma}_k=\bsbGamma_k$ if $\|\bsbGamma_k\|_2$ is among the $m$ largest norms in the set of   $\{\|\bsbGamma_k\|_2: k=(i, j), 1\leq i < j \leq p\}$,  and $\tilde{\bsbGamma}_k=\bsb{0}$ otherwise. This iterative quantile screening was proposed in \cite{SheGLMTISP} and has found successful applications in group selection, rank reduction, and network screening \cite{SheSpec,SheTISPMat,sel-rrr,s2,sigmoid}.

An equivalent  way to perform the multivariate quantile thresholding $\vec{\Theta}^\# (\bsb{\Gamma}; m)$ for any  $\bsbGamma=[\bsbB, \phi\bsbOmega]$ is based on the JAG. First,   compute  the JAG matrix $\bsb{{C}}$ by \eqref{eq:defineC} explicitly for all $i\neq j$, and set all its diagonal entries to be $0$. Then perform  \textit{elementwise} hard-thresholding on  $\bsbC$  with the threshold set as  the $(2m+1)$th largest element in $\bsbC$. Finally, zero out `small' entries in $\bsbGamma$ or $(\bsbB, \bsbOmega)$:  for  any $i\neq j$, set
$ b_{ij} = \omega_{ij} = 0 ,\text{ if } c_{ij}=0$.  
See Algorithm \ref{alg:JAG} for more details. From~\cite{SheSpec}, we can similarly show that the iterative quantile thresholding converges and leads to a local minimum of  the following $\ell_0$-constrained problem:
\begin{equation}
\label{eq:l0_problem}
\begin{gathered} \min_{\bsbB,\bsb{\Omega}\succ 0}
L(\bsbB,\bsb{\Omega}) \quad
\text{ s.t. } \|\bsb{C}\|_{0}^{\mbox{\small off}} \leq q (p^2-p),
\end{gathered}
\end{equation}
where $\|\bsb{C}\|_0^{\mbox{\small off}}$ denotes the number of nonzero off-diagonal entries in  $\bsb{C}$, and $q$ ($0< q \leq 1$),  called the \emph{quantile parameter}, puts an upper bound on the sparsity level of the network. It can be  customized by the user based on the belief of how sparse the network could be. Prior knowledge or specific application needs can be incorporated. Thus this upper bound is usually not difficult to specify  in sparse network learning.

In the generalized linear model setting, the proposed iterative multivariate thresholding procedure  is
guaranteed to converge with a simple analytical expression for the
step size $\alpha^l$~\cite{SheGLMTISP,SheSpec}. However,  in  our dynamical network which has
both $\bsbB$ and $\bsb{\Omega}$ unknown,  there seems to be no simple formula
for  the step size in \eqref{eq:l0_problem}. The constraint cone $\bsbOmega\succ 0$ increases the difficulty in deriving a universal  step size.
We propose a {simple} but {effective} asynchronous Armijo-type (denoted as \textbf{AA}) line search approach in the next subsection, which guarantees a convergent solution with $\bsbOmega\succ 0$ automatically satisfied.

\subsection{The AA line search and the GIST algorithm}
\label{sec:Atype}
The basic idea of the Armijo-type line search, when restricting to problem \eqref{eq:l0_problem}, is to select a step size along the descent direction that satisfies the Armijo rule~\cite{Armi66a}:
\begin{equation}
\bar L(\bsb{\Gamma}^{l+1}) \leq \bar L(\bsb{\Gamma}^l) + c_1 \tr\{
(\bsb{\Gamma}^{l+1} - \bsb{\Gamma}^l )^\tran\bsb{G}^{l}\}.
\label{eq:armijo}
\end{equation}
If the condition is satisfied, we accept $\bsb{\Gamma}^{l+1}$ and
carry on to the next iteration; otherwise,  decrease $\alpha^l$ and try
the new update in \eqref{eq:quantile_th} till either the condition is satisfied or $\alpha^l$
becomes smaller than a threshold $c_2$. The values of $c_1, c_2$ can be
set to, for instance,  $c_1 = 10^{-4}$ and $c_2 = 10^{-6}$. At each iteration we initialize
$\alpha^l$ as $1$ and decrease $\alpha^l$ according to $\alpha^l \leftarrow \alpha^l/10$ if the condition \eqref{eq:armijo}
is not satisfied.

Empirical studies show that $\bsbG_B$ and $\bsbG_\Omega$ usually have different orders of magnitude and so using the same step size for updating $\bsbGamma_B$ and $\bsbGamma_\Omega$ may be suboptimal. (In fact, with only one step size parameter,  it is often difficult  to find an $\alpha^l$ satisfying \eqref{eq:armijo}, and  the algorithm  converges slowly.) Therefore, we propose using different step sizes for $\bsbG_B$ and $\bsbG_\Omega$.
This can be    implemented by  updating the $B$-component and  the $\Omega$-component
asynchronously in computing $\bsbC$.
To be more specific, we
modify \eqref{eq:derivative} as
\begin{equation}
\bsb{G}^l = \begin{cases}
[\bsb{G}_B^l, \bsb{0}] & \text{ if $l$ is odd} \\
[\bsb{0}, \phi^{-1}\bsb{G}_{\Omega}^l] & \text{ if $l$ is even.}
\end{cases}
\end{equation}
The AA line search can implicitly guarantee
the positive definiteness of $\bsbOmega$. If we set
$\log |\bsb{\Omega}| = -\infty$ for any
$\bsb{\Omega}$ not positive definite,  then such  $\bsbOmega$'s will naturally be rejected by  \eqref{eq:armijo}. The same treatment  has been used in Gaussian graph learning, see, e.g.,~\cite{Duchi08_BlockSparsInvCov}.

The  final form of our  \textit{graph iterative screening via thresholding} (GIST) is proposed in Algorithm \ref{alg:JAG},   under the assumption that the data $\bsbX$ has been centered and normalized column-wise,  $\bsbY$ has been centered, and  $\bsb{\Sigma}_{XX}\eqdef\bsb{X}^\tran\bsb{X}$ and $\bsb{\Sigma}_{XY}\eqdef\bsb{X}^\tran\bsb{Y}$.


\begin{algorithm}
{\footnotesize{
\caption{GIST for JAG screening}
\label{alg:JAG}
\begin{algorithmic} 

\REQUIRE Data matrices $\boldsymbol{X}, \boldsymbol{Y},
\bsb{\Sigma}_{XX}, \bsb{\Sigma}_{XY}$; quantile $q$; parameters for AA
line search $c_1$, $c_2$; maximum iteration number $M$; error
tolerance $\xi_C$; $\phi$: weight parameter in JAG construction; initial estimates $\bsbB^0, \bsb{\Omega}^0$. \\

\STATE \emph{1) Initialization:}  $f \gets L(\bsbB^0, \bsb{\Omega}^0
); l\gets 0$;

\emph{2) Perform the AA line search:}

\REPEAT

\STATE $\alpha^l\gets 1$;

\REPEAT

\STATE \emph{2.1) }Update $\bsbB$ and $\bsbOmega$:

 \IF{$l$ is odd} \STATE
$\bsb{G}_B^l \leftarrow (\bsb{\Sigma}_{XX}\bsbB^l-\bsb{\Sigma}_{XY})\bsb{\Omega}^l$;
$\bsbB^{l+1} \leftarrow \bsbB^l-\alpha^l \bsb{G}_B^l$;
$\bsbOmega^{l+1} \gets \bsbOmega^l$;
$\Delta^l \leftarrow \tr\{(\bsbB^{l+1}-\bsbB^{l})^\tran\bsb{G}_B^l\}$;

\ELSE \STATE
$\bsb{G}_{\Omega}^l \leftarrow\frac{1}{2}(\bsb{Y}-\bsb{X}\bsbB^l)^\tran(\bsb{Y}-\bsb{X}\bsbB^l)- \frac{n}{2} (\bsb{\Omega}^l)^{-1}$;
 $\bsb{\Omega}^{l+1} \leftarrow \bsb{\Omega}^l-\alpha^l \bsb{G}_{\Omega}^l$;
 $\bsbB^{l+1}\gets \bsbB^l$;
 $\Delta^l \leftarrow \tr \{(\bsbOmega^{l+1}-\bsbOmega^{l})^\tran\bsb{G}_\Omega^l \}$;

\ENDIF

\STATE \emph{2.2)} $\bsbC^{l+1} \gets [c_{ij}^{l+1}]$, where $c_{ii}^{l+1}=0$ ($1\leq i \leq p$),    $c_{ij}^{l+1}= c_{ji}^{l+1} = \sqrt{(b_{ij}^{l+1})^2+(b_{ji}^{l+1})^2+2\phi^2 (\omega_{ij}^{l+1})^2}, \forall i, j: i \neq  j$;

\STATE \emph{2.3)}  $\lambda^{l+1} \gets \text{the }(2\lceil q(p^2 -p)\rceil +1)\text{th largest element in } \bsb{C}^{l+1}$;

\STATE \emph{2.4)} $\bsb{S} \gets \mbox{sgn}(\Theta_H(\bsbC^{l+1}; \lambda^{l+1}) + \bsb{I})$, where $\mbox{sgn}$ is the elementwise sign function and $\Theta_H$ performs  {elementwise} hard-thresholding;

\STATE \emph{2.5)} $\bsbB^{l+1}\gets \bsbB^{l+1} \circ \bsb{S}, \bsbOmega^{l+1}\gets \bsbOmega^{l+1} \circ \bsb{S}$, where ``$\circ$'' denotes the Hadamard product;

\STATE $f^{l+1} \leftarrow L(\bsbB^{l+1},\bsb{\Omega}^{l+1} )$; $\alpha^l \leftarrow
\alpha^l/10$;

\UNTIL{$f^{l+1} \leq f^l+c_1 \Delta^l$ or $\alpha^l\leq c_2$ }

\STATE $l \leftarrow l+1$;

\UNTIL{$|f^l-f^{l-1}| \leq \xi$ or $l \geq M$ or the pattern of
$\bsb{C}^l$ stops changing}
\ENSURE JAG estimate $\bsb{\hat{C}}=\bsbC^{l}$ and its screening pattern $\{(i,j):\hat{c}_{ij}\neq 0\}$.
\end{algorithmic}
}}
\end{algorithm}


The GIST algorithm is very simple to implement and runs efficiently. If the purpose is to get the convergent sparsity pattern  instead of the precise estimate, one can terminate the algorithm  as long as the sign of the
iterates stabilizes---usually within 50 steps.
Even for a network with 500 nodes, GIST takes less than 1 second.

\subsection{Robust JAG decomposition via spectral clustering}
\label{sec:decomposition}
 Nowadays, a great challenge in modern network analysis comes from big data,
which makes many methods computationally infeasible.
Fortunately, very large networks often demonstrate subnetwork structures and thus one can decompose the network in an early stage, and then apply complex learning algorithms to each subnetwork individually. Similar ideas have appeared in Gaussian graph learning~\cite{Witten2011, Rahul2012}, where a simple {one-step} thresholding is applied to the sample covariance matrix to pre-determine if the associated concentration matrix estimate is decomposable, referred to as the \textit{block diagonal screening rule} (BDSR). Yet  it  ignores the first-order statistical structure of our dynamical model \eqref{eq:model}, and  the resulting CDG may not reliably capture the network topology. See  Section~\ref{sec:SP500} for some experiments.

We propose  decomposing the whole network based on the GIST estimate. For example, after getting $\bsb{\hat{C}}$ ,we can apply the \textit{Dulmage-Mendelsohn Decomposition}~\cite{dulmage1958coverings} to detect if there exists an exact block diagonal form of $\bsb{\hat{C}}$. However, the noise contamination makes perfect decomposition  seldom possible. Therefore, we  treat $\bsb{\hat{C}}$ as a similarity matrix where the ``association strength'' $c_{ij}$  indicates how close node $i$ and node $j$ are, and so pursuing an approximate block diagonal form is now identified as a node \textbf{clustering} problem.
Specifically, we apply \textit{Spectral Clustering} to $\bsb{\hat{C}}$ %
to obtain  a  {robust} JAG decomposition. Refer to \cite{Von2006tutorial_spectral} for a comprehensive introduction,
and \cite{von2008consistency} for its ability in suppressing the  noise.  There are many effective  ways to determine the number of clusters~\cite{fraley1998many, tibshirani2001estimating, sugar2003finding}.

Unlike    \cite{Witten2011} and \cite{Rahul2012},  our JAG decomposition  does not rely on  setting  $q$  low in \eqref{eq:l0_problem}     to  yield  subnetworks. An over-sparse estimate may be problematic in estimation or structure identification.
The philosophy is   different  from that of the BDSR. In fact, BDSR is purely computational---it pre-determines, for each $\lambda$, if the associated graph estimate is perfectly decomposable or not.    To ensure decomposability on noisy data, one  tends to specify overtly high sparsity levels  to obtain subnetworks---see, e.g., Section 4 in \cite{Rahul2012} and our data example in Section \ref{sec:SP500}. This may remove genuine  connections. Therefore,  the resultant  decomposition    could be misleading, and excessive  bias  may be  incurred  in estimation  (cf. \cite{Zhao2012_huge}).
Our JAG decomposition can deal with noise effectively and is much more robust in  this sense.

If the network is decomposable (or approximately so),  system \eqref{eq:model}
can be re-written as
$
\bsb{x}_t^i=\bsbA_{ii}\boldsymbol{x}_{t-1}^i+\bsb{\epsilon}_t^i,
\quad \bsb{\epsilon}_t^i \sim \mathcal{N}(\bsb{0},
\bsb{\Sigma}_{ii}), 
$
for $i=1,\cdots,d$, where $d$ is the number of subnetworks and $\bsb{x}^i$ corresponds to the nodes that
belong to the $i$th subnetwork. We can thereby conduct
fine estimation of $\bsbB_{ii}$ and $\bsbOmega_{ii}$ for each
subnetwork (in a possibly parallel fashion).
There are two ways to use the GIST screening outcome. If each subnetwork is of relatively small size such that fine learning algorithms can be applied smoothly, one can drop the constraints in \eqref{eq:stage2}. In this case,  $\bsb{\hat{C}}$ is only used to reveal the block decomposition structure. Alternatively, one can enforce all within-block sparsity constraints (determined  by $\bsb{\hat{C}}_{ii}$) in  sub-network learning. The latter is  usually faster, but  when the value of $q$ is set too low, one should caution against such a manner.

\section{Fine $(\bsbB, \bsbOmega)$ learning}
\label{sec:stage2}
In this stage of JGSE we   perform  fine estimation of the graph matrices. Recall the optimization problem to take advantage of   the JAG screening pattern given by Stage 1$$
\begin{gathered}
\min_{\bsbB,\bsb{\Omega}\succ 0}
L(\bsbB,\bsb{\Omega})+P_B(\bsbB;\lambda_B)+P_{\Omega}(\bsb{\Omega};\lambda_{\Omega})\\
\text{ s.t. } E_B \subset E_{\hat{C}}, E_{\Omega} \subset
E_{\hat{C}}.
\end{gathered} \nonumber
$$
Sometimes the screening constraints may   be dropped. In either case, we can state the  optimization problem as instances of
\begin{equation}
\min_{\bsbB,\bsb{\Omega}\succ 0}
L(\bsbB,\bsb{\Omega})+P_B(\bsbB;\bsb{\Lambda}_B)+P_{\Omega}(\bsb{\Omega};\bsb{\Lambda}_{\Omega}),\\
\label{eq:stage2_2}
\end{equation}
where $\bsb{\Lambda}_B=[\lambda_{B,ij}]$ and $\bsb{\Lambda}_{\Omega}=[\lambda_{\Omega,ij}]$ are regularization parameter matrices. Indeed, to enforce the screening constraints, we can set
$
\lambda_{B,ij} = \infty  \text{ if } \hat{c}_{ij}=0$ and $
\lambda_B  \text{ otherwise,}$ and
$\lambda_{\Omega,ij} =  \infty  \text{ if } \hat{c}_{ij}=0$, and
$\lambda_\Omega \text{ otherwise. }$

To  solve for $\bsbB$ with $\bsb{\Omega}$ held fixed,
it suffices to study
\begin{align}
\min_{\bsbB} f_B(\bsb{B};\bsb{\Lambda}_B)=  \frac{1}{2}\tr\{(\bsb{Y}-\bsb{X}\bsbB)
\bsb{\hat{\Omega}}(\bsb{Y}-\bsb{X}\bsbB)^\tran\}+P_B(\bsbB;\bsb{\Lambda}_B),
\label{eq:B_Ofixed} 
\end{align}
With $\bsbB$ fixed,  the  problem of interest reduces  to
\begin{equation}
\min_{\bsb{\Omega}\succ 0}
\frac{1}{2}\tr\{(\bsb{Y}-\bsb{X}\bsb{\hat{B}})
\bsb{\Omega}(\bsb{Y}-\bsb{X}\bsb{\hat{B}})^\tran\}-\frac{n}{2}\log
|\bsb{\Omega}|+P_{\Omega}(\bsb{\Omega};\bsb{\Lambda}_{\Omega}).
\label{eq:O_Bfixed} 
\end{equation}
Fortunately, the optimization  still falls into the framework described in Section~\ref{sec:group_theta}. We introduce the  \textit{fine learning of graphs} (FLOG) algorithm as follows.  For simplicity, suppose $P_B$ and $P_\Omega$ are $\ell_1$ penalties.
Algorithm \ref{alg:JAG}  can be adapted to  the $\bsbB$-optimization \eqref{eq:B_Ofixed} (with $\bsb{\Omega}$ fixed at its current
estimate $\bsb{\hat{\Omega}}$, and under the initialization  $l=0$, $\alpha=1$ and $\bsbB^l=\bsb{\hat{\bsbB}}$):

{{
\begin{algorithmic}
\STATE $\bsb{G}_B^l \gets
(\bsb{\Sigma}_{XX}\bsbB^l-\bsb{\Sigma}_{XY})\bsb{\hat{\Omega}}$;
\REPEAT

\STATE $\bsbB^{l+1} \gets \Theta_S(\bsbB^l-\alpha\bsb{G}_B^l;
\bsb{\Lambda}_B)$;

\STATE $\alpha \leftarrow
\alpha/10$;

\UNTIL{$f^{l+1}_B \leq f^l_B + c_1 \tr\{
(\bsbB^{l+1}-\bsbB^l)^\tran(\bsb{G}_B^l +  \mbox{sgn}(\bsb{\Lambda}_B\circ \bsbB^l))\}$ or $\alpha \leq c_2$ (by convention $\infty\cdot 0=0$)}

\STATE $l\leftarrow l+1$;

\end{algorithmic}
}}
%
Experimentation shows that the line search performance is not sensitive to the values of $c_1$ and $c_2$; we simply set $c_1=10^{-4}$ and $c_2=10^{-6}$, following~\cite{nocedal2006numerical}.
As for  the $\bsbOmega$-optimization  \eqref{eq:O_Bfixed}, this is just the  Gaussian graph learning problem with the sample covariance matrix given by $\frac{1}{n}(\bsb{Y}-\bsb{X}\bsb{\hat{B}})^\tran(\bsb{Y}-\bsb{X}\bsb{\hat{B}})$. The popular
graphical lasso ~\cite{Friedman08_graphLasso} can be used.

\textbf{Some related works.} The MRCE algorithm solves a similar  fine learning problem to \eqref{eq:stage2_2}, but  there exist no screening constraints. \citet{lee2012simultaneous} generalized MRCE to handle weighted penalties. Both algorithms use cyclical coordinate descent in the $\bsbB$-optimization step, which has a
worst case cost $O(p^4)$~\cite{Rothman2010_SparseBsparseTheta}. In contrast, the proposed $\bsbB$-update in FLOG has complexity
 $O(p^3)$, which comes from
the $p \times p$ matrix multiplication for computing the gradient
$\bsb{G}_B$.
Experiments show that FLOG is more efficient than MRCE under the same setting of error tolerance and maximum iteration numbers. 

With FLOG introduced, the two-stage JGSE learning framework is complete. We  point out that although FLOG is more efficient and scalable than MRCE, the main contribution of JGSE lies in Stage 1 which reduces the problem size and search space for fine estimation.

\section{Experiments on synthetic data}
\label{sec:experiment}
In this section, we show the performance of  GIST and FLOG  in the JGSE network learning using synthetic data.
\subsection{Identification and estimation accuracy}
\label{sec:experiment_accuracy}
We compare the proposed JGSE with some relevant methods in the literature:
\begin{compactitem}
\item sGTG estimates the sparse $\bsbB$ only, assuming $\bsbOmega\propto \bsb{I}$. It is implemented using the coordinate descent algorithm~\cite{friedman2007pathwise}.
\item sCDG estimates the sparse concentration matrix $\bsbOmega$, after centering the data. It is implemented using the graphical lasso \cite{Friedman08_graphLasso}.
\item MRCE~\cite{Rothman2010_SparseBsparseTheta} jointly estimates $\bsbB$ and $\bsbOmega$ subject to separate penalties, and its implementation is given by the R package ``MRCE''.
\end{compactitem}
In all the methods, the $\ell_1$ penalty function is used for $P_B$/$P_{\Omega}$.
Experiments are performed for  the following  networks with different sizes and topologies.
\begin{compactitem}
\item Example 1: $p=40, n=100.$ The network consists of two equally sized subnetworks.
\item Example 2: $p=80, n=200.$ The network consists of three subnetworks of sizes 40, 20,  20.
\item Example 3: $p=160, n=300.$ The network consists of four equally sized subnetworks.
\item Example 4: $p=20, n=50$,  $\bsbOmega=\bsb{I}$. $\bsbB$ has no subnetwork structure.
\item Example 5: $p=20, n=50$, $\bsbB = \bsb{0}$.  $\bsbOmega$ is non-diagonal, and shows no subnetwork structure.
\end{compactitem}
The identification accuracy is measured by the true positive rate $\mbox{TPR} =
\frac{\#\{(i,j): \hat{c}_{ij}\neq 0,  c_{ij}\neq
0\}}{\#\{(i,j): c_{ij}\neq 0\}}$ and false positive rate
$\mbox{FPR} =
\frac{\#\{(i,j): \hat{c}_{ij}\neq 0,  c_{ij}=
0\}}{\#\{(i,j): c_{ij}= 0\}}$.
In Examples 1-4, the estimation accuracy is measured by the model error $ME_{\bsbB} =\tr\{(\boldsymbol{\hat{B}}-
\boldsymbol{B})^\tran\boldsymbol{\Sigma}_{XX}(\boldsymbol{\hat{B}}-
\boldsymbol{B})\}$  \cite{Yuan07}. In Example 5,  only $\bsbOmega$ is estimated, and the  accuracy is measured by $ME_{\bsbOmega}=\|\hat{\bsbOmega}-
\boldsymbol{\Omega}\|_F^2$.

 In each of the settings, the number of unknown variables is much larger than the number of observations, e.g., $p^2+p(p+1)/2=9,640 \gg 200$ in Example 2. The diagonal blocks of $\bsbB$ and $\bsbOmega$ are all sparse random matrices generated independently, following the scheme in \cite{Yuan07}. 
The data observations  are then generated  from the multivariate time series model \eqref{eq:model}.
We repeat the synthetic data experiment in each setting for  50 times and summarize the performance of an algorithm as follows.  
 Mean  TPR and FPR   are reported. The distribution of ME appears non-Gaussian and multimodal; for robustness and stability,   the 25\% trimmed-mean of
model errors from multiple runs is reported.   The algorithms include   sGTG, sCDG, MRCE, FLOG$^w$ and JGSE.
FLOG$^w$ is to make a comparison with MRCE, and denotes FLOG applied to the whole network, i.e., running the second stage algorithm of JGSE without the first stage JAG screening.  (We point out however that this is  \textit{not} the recommended way of network estimation in the paper; our proposed JGSE applies FLOG \emph{after} GIST screening.)  The JAG weight parameter $\phi$ is taken to be 1 throughout all experiments.
In Examples 1-3, spectral clustering is called after running sGTG,  sCDG, MRCE, and FLOG$^w$,    because of the existence of subnetworks.  All regularization parameters are chosen by minimizing the model validation error, evaluated on 1,000 validating samples independently generated in addition to the training data.
We set the value of $q$  to be $0.3$, which is  large enough for screening. (Tuning the quantile parameter showed no observable difference; its robustness is also seen in Section \ref{sec:experiment_decomposition}.)
Table~\ref{tab:identification}  shows the results.
\begin{table*}
  \centering
  {\small{
  \caption{\small{Method comparison in terms of  true positive rate (TPR), false positive rate (FPR) and model error (ME).}}\label{tab:identification}
\begin{tabular}{r|c c c c c}    \hline
     &        Example 1       &         Example 2     &       Example 3            &       Example 4        &       Example 5       \\
& (TPR,  FPR),  ME & (TPR, FPR), ME & (TPR, FPR), ME & (TPR, FPR), ME & (TPR, FPR), ME \\
\hline
sGTG & (24\%, 4\%), 1947.8 & (16\%, 2\%), 6404.6 & (12\%, 1\%), 21360.3 & (89\%, 20\%), 53.5 & (29\%, 6\%), 182.1\\
\hline
sCDG & (47\%, 28\%),  N/A & (32\%, 17\%),  N/A & (26\%, 10\%),  N/A & (70\%, 46\%),  N/A &  (93\%, 38\%),  2.8\\
\hline
MRCE & (63\%, 13\%), 106.6 & (61\%, 8\%), 187.9 & Infeasible   & (88\%, 29\%), 98.6 & (76\%, 20\%),  7.3\\
FLOG$^{w}$ & (83\%, 25\%), 100.9 & (91\%, 21\%), 166.8 & (88\%, 13\%), 635.1 & (87\%, 21\%), 68.8 & (88\%, 45\%),  5.6\\
\hline
JGSE & (91\%, 28\%), \textbf{74.6} & (95\%, 23\%), \textbf{140.8} & (95\%, 14\%), \textbf{549.4} & (85\%, 10\%), \textbf{53.6} & (87\%, 44\%),  \textbf{5.6} \\
\hline
\end{tabular}
 }}
\end{table*}

In Examples 1-3, both sGTG and sCDG suffer from over-simplified model assumptions and fail to identify network connections  accurately.
Indeed, we frequently observe  that the conditional dependence graph from sCDG is  not sufficiently sparse. It seems that sCDG tries to rephrase  first-order dynamics as node correlations  and consequently results in a dense second-order topology. sGTG  shows lowest  TPR values and misses many true connections, which is a sign of  over-shrinkage. Not surprisingly, in the two degenerate cases,  sGTG behaves well  in Example 4 (because $\bsbOmega=\bsb{I}$), and similarly, sCDG  does a good job in Example 5 where $\bsbB=\bsb{0}$.

MRCE estimates both  first-order and second-order statistics and achieves much lower error rates than   sGTG (except in Example 4). However, MRCE is quite computationally expensive and may be infeasible for large-scale problems. In Example 2, it took MRCE around 40 minutes to run a single experiment. In Example 3, MRCE became computationally intractable.
FLOG$^w$ did not show  such  computational limitations there. The two algorithm designs resulted in different estimates. (Recall that the objective criterion is nonconvex.) MRCE is less accurate in general.

The complete JGSE learning is even more efficient,  owing to the first stage GIST for robust JAG screening and decomposition.  More importantly, JGSE shows remarkable improvements in    estimation   in almost all  cases. (The only exception is Example 5, where JGSE has  comparable performance to FLOG$^w$.) These positive  results validate  the power of  GIST in removing lots of unnecessary edges and reducing the search space for topology identification.    In all, our two-stage JGSE (GIST+FLOG) successfully beats   the existing joint graph learning method MRCE.


\subsection{GIST in Decomposition }
\label{sec:experiment_decomposition}
In this subsection, we examine the performance of   GIST in network decomposition.
The rand index (RI)~\cite{Rand_randIndex} is used for evaluation. It is obtained by comparing the memberships of each pair of nodes assigned by an algorithm with
 the true memberships. If a pair coming from
 the same cluster are assigned to a single cluster, it is defined as a true
 positive ($TP$); if a pair coming from different clusters are
 assigned to different clusters, it is defined as a true negative
 ($TN$); $FN$ and $FP$ are defined similarly.
 Then RI is defined as $(TP+TN)/(TP+TN+FP+FN)$.

We fix a small sample size $n=30$ and vary $p$ in this experiment. The time series data are again generated according to the multivariate auto-regression \eqref{eq:model}. All networks consist of two equally-sized subnetworks; each diagonal block of $\bsbB$ is generated as a random sparse
 matrix, and  each diagonal block of $\bsbSigma$ has  diagonal elements  1 and all  off-diagonal
 elements  $0.5$. 
Each experiment is repeated 50 times.

\begin{figure*}
  \centering
  \subfloat[\footnotesize RI versus quantile ($p=500$)]{\label{fig:RI_q}\includegraphics[width=0.4\textwidth]{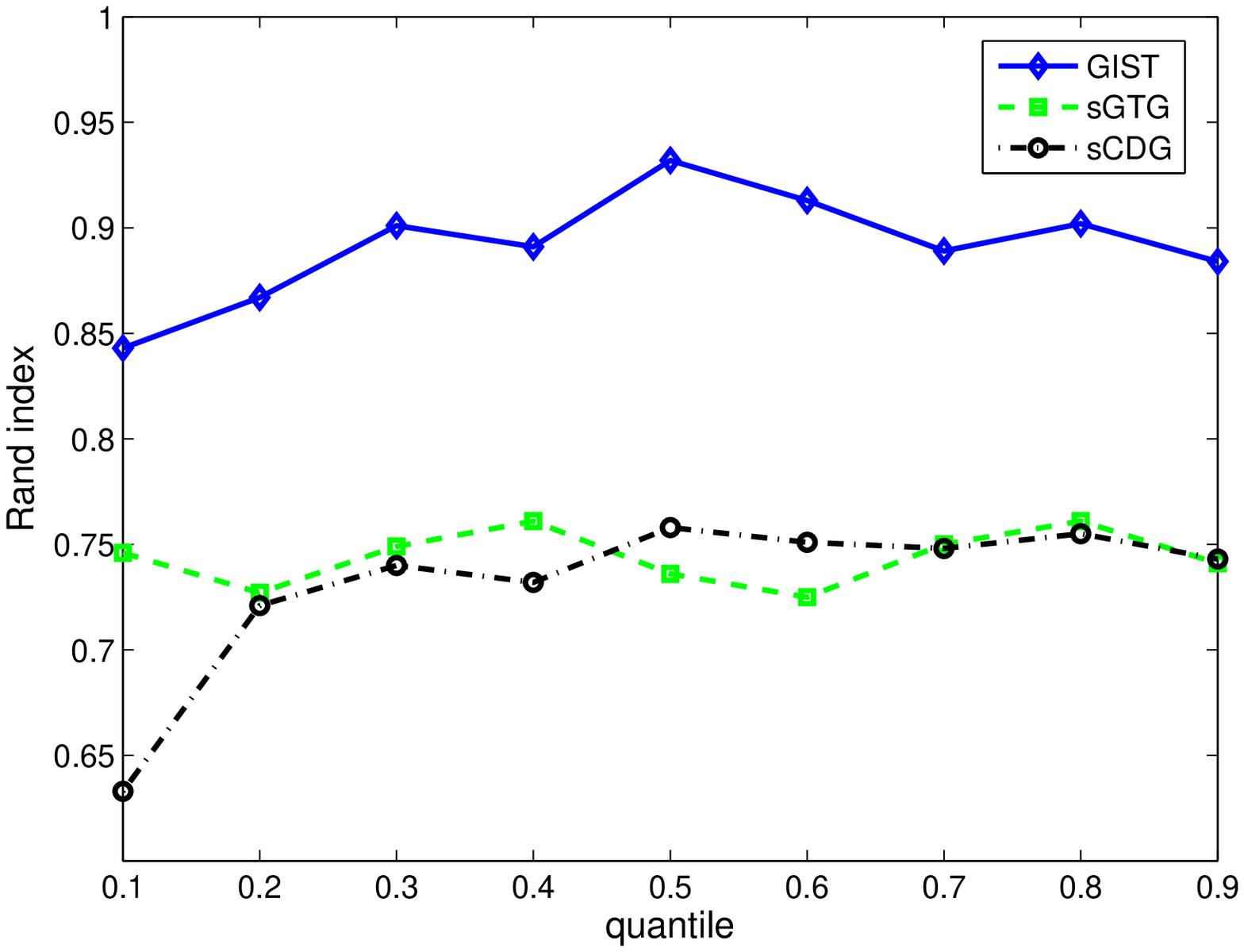}}
  \subfloat[\footnotesize RI versus network size ($q=0.5$)]{\label{fig:RI_p}\includegraphics[width=0.4\textwidth]{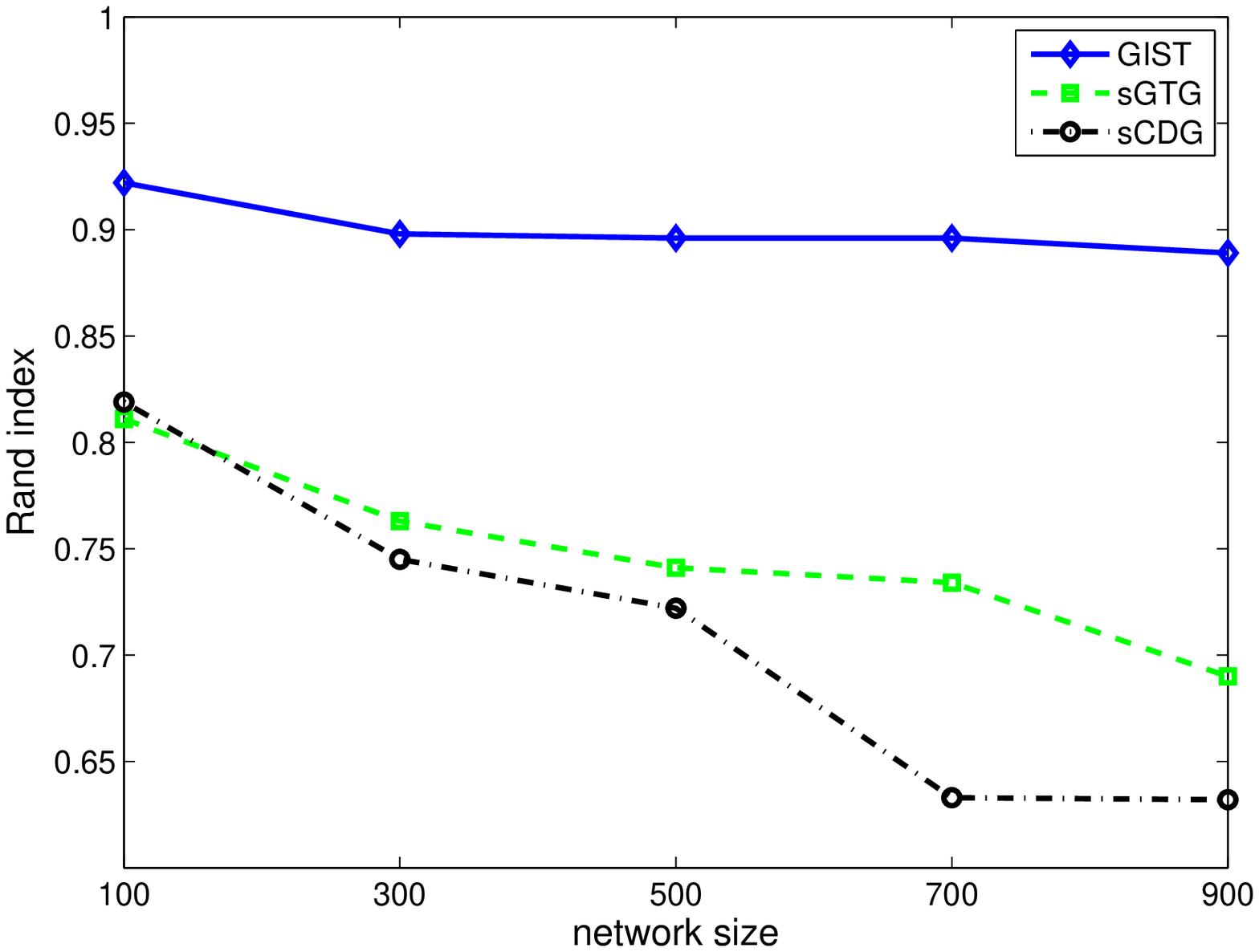}}
  ~ 
  \caption{\footnotesize Rand index comparison on simulated networks consisting of two equally-sized subnetworks.} \label{fig:clusteringErr}
\end{figure*}

Given any network data, we apply  GIST to  obtain a JAG estimate and perform robust  decomposition (cf. Section~\ref{sec:decomposition}).   The decompositions of  sole GTG (assuming $\bsb{\hat{\Omega}}_{sGTG}=\bsb{I}$) and  sole CDG (assuming $\bsb{\hat{B}}_{sCDG}=\bsb{0}$ after centering the data) are obtained as well.
All decompositions  are  via spectral clustering. Although GIST considers a more complex model, because of its screening nature, it runs efficiently.
The mean RI results are shown in Figure~\ref{fig:clusteringErr}.

In all the settings, GIST achieves more reliable decomposition and outperforms sGTG and sCDG by a large margin.   This shows  that  the network decomposition based on the joint graph is trustworthy.
Moreover, its performance is rather \textbf{insensitive} to the choice of the quantile parameter $q$ as long as $qp^2$ bounds the true network cardinality.  This offers great ease in practice.

GIST is also superfast: for any network in the experiments, it just takes a few seconds to obtain the graphical screening pattern or subnetwork structure. A more comprehensive computational cost investigation is given in the next subsection.

%
\subsection{Computation time reduction via graph screening}
\label{sec:experiment_time}
Now we study  how much computational cost can be saved by applying  GIST before fine learning. All simulated networks
consist of multiple equally-sized subnetworks, with the total number of nodes denoted by $p$ and the number of nodes in  each subnetwork denoted by $p_s$. The diagonal blocks of $\bsbB$ and $\bsbOmega$ are generated in the same manner as in Section~\ref{sec:experiment_decomposition}. Let $T_{JGSE}$ be
the total computation time of
 JGSE learning  (``GIST+FLOG''),
and $T_{FLOG}^{(w)}$
be the computation time  by applying
FLOG {directly} to the \textit{whole}
network without graph screening or decomposition. (We did not include MRCE in the comparison because it is extremely slow for large data). Solution paths of $\bsbB$ and $\bsbOmega$ are computed for  a grid of values for $(\lambda_B, \lambda_\Omega)$ that covers various  sparsity patterns. The quantile parameter is set as 0.3 in GIST. We report the ratios $T_{JGSE}/T_{FLOG}^{(w)}$ for different combinations of ($p$, $p_s$) in Table~\ref{tab:time}, where $n=100$ in all  experiments.
\begin{table*}
  \centering
  {
  {
  \caption{
  Computation time reduction offered by JGSE on simulated networks, with  $p$ denoting the total number of nodes and $p_s$  the number of nodes for each  subnetwork.}\label{tab:time}
\begin{tabular}{c|cccccc}
  \hline
  $(p, p_s)$   & (500,250) & (500,100) & (500,50) & (1000,500) & (1000,200) & (1000,100)
\\ \hline
  $T_{JGSE}/T_{FLOG}^{(w)}$ & 0.427   & 0.161   & 0.133  & 0.404 & 0.148 & 0.112 \\
  \hline
\end{tabular}
}}
\end{table*}

Table~\ref{tab:time} shows that at least  half of the running time can be reduced when the network is decomposable.
The larger the ratio $p/p_s$ is,
the more computational cost can be saved. We conducted the experiment  on a  PC, but if parallel computing resources are available, the computational efficiency can be further boosted.
The network decomposition technique makes an otherwise computationally expensive or even infeasible problem  much easier to solve. 

\section{Applications}
\label{sec:application}
In this section, we  analyze real data from $S\&P$ 500 and NASDAQ-100 stock using JGSE.

 \subsection{S\&P 500}
 \label{sec:SP500}
This dataset keeps a record of  the closing
prices of the $S\&P$ 500 stocks from Jan. 1, 2003 to Jan. 1,
2008. It consists of 1258 samples for
452 stocks.
The data have been preprocessed by  taking logarithm and differencing transformations~\cite{Zhao2012_huge}.

\begin{figure*}
\centering
  \subfloat[\footnotesize JAG decomposition ($q=0.1$). ]{\label{fig:jag_cluster} 
  \includegraphics[width=0.4\textwidth]{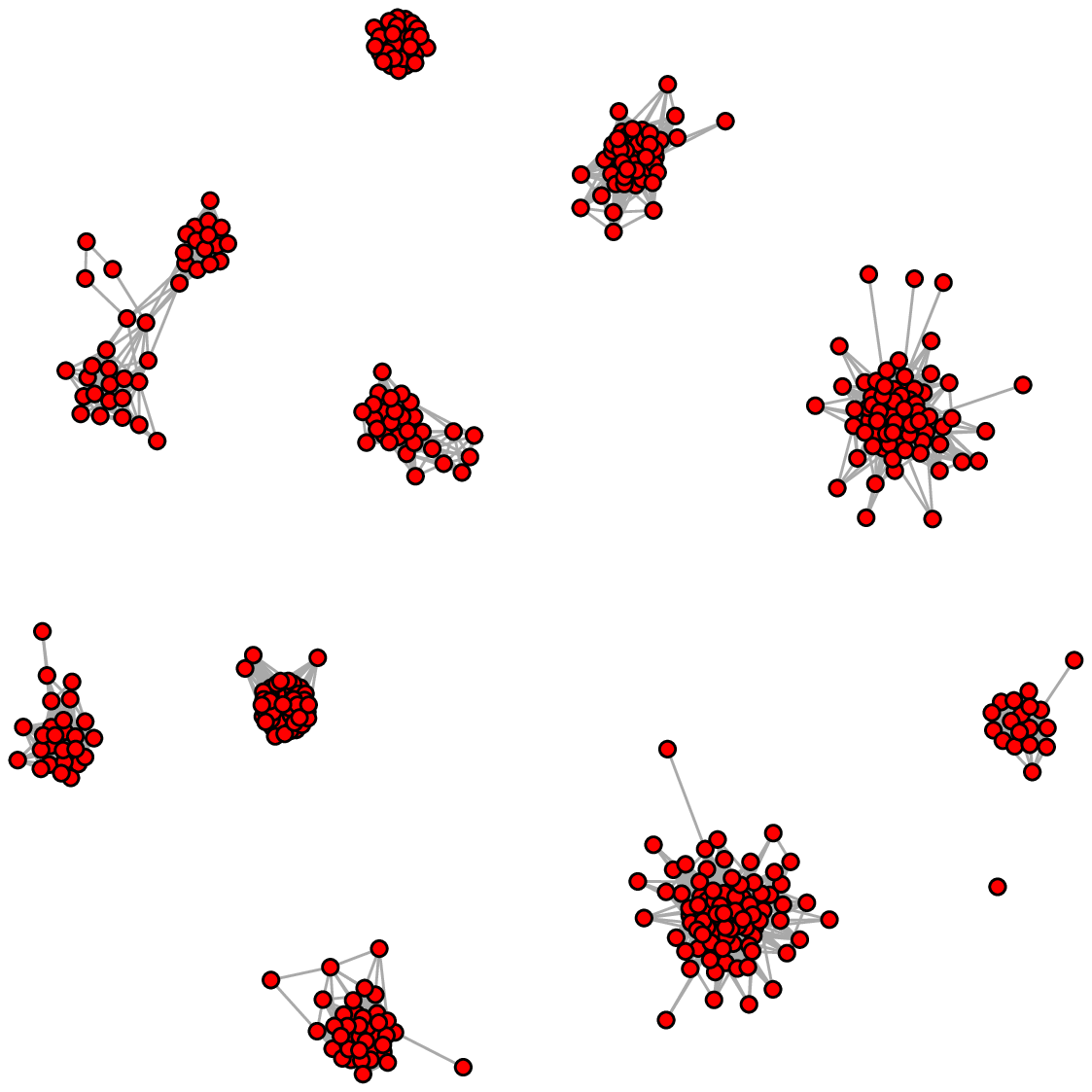}}  
  \subfloat[\footnotesize Rand index comparison. ]{\label{fig:randIndexSP} 
   \includegraphics[width=0.4\textwidth]{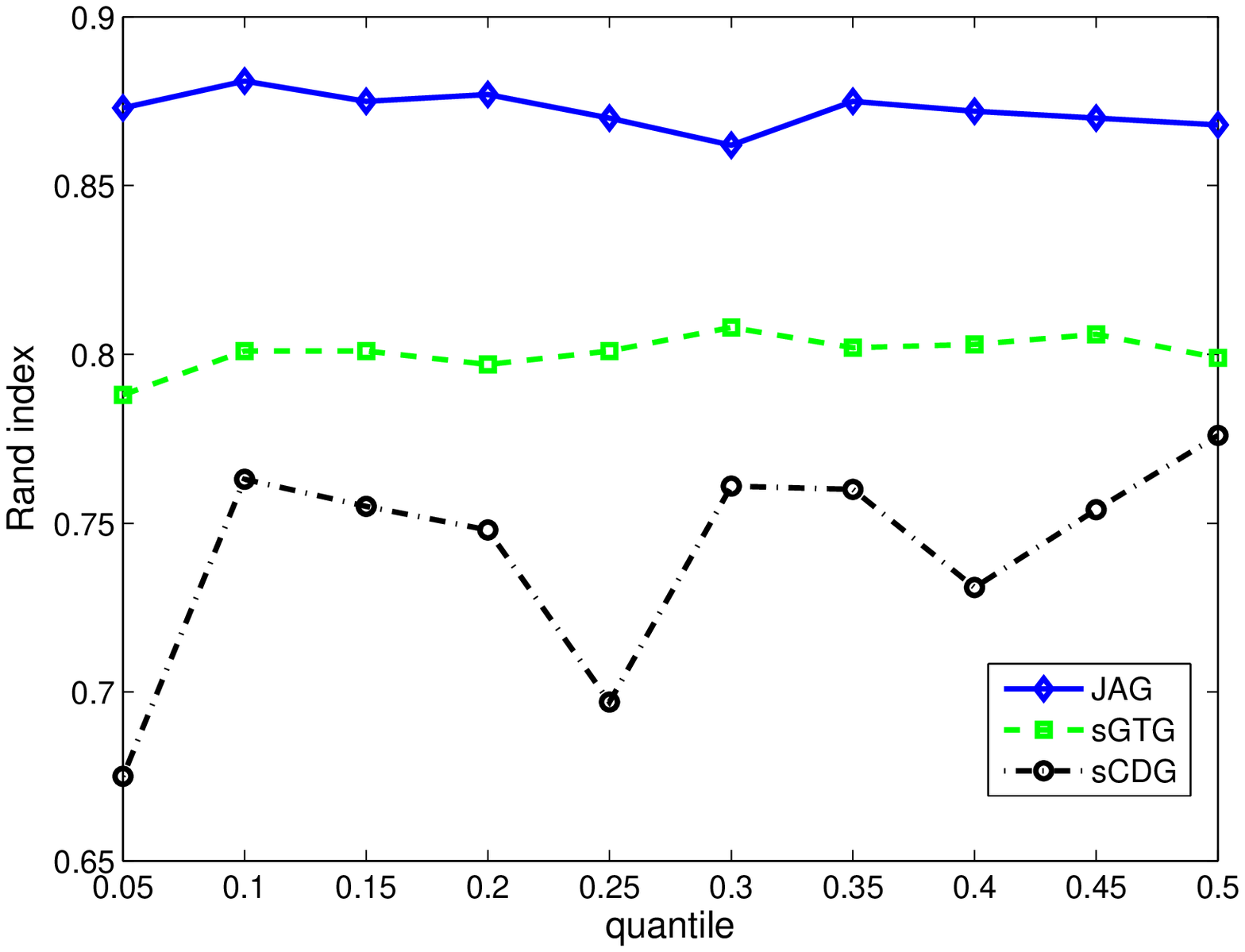}}
  \caption{\footnotesize GIST on $S\&P$ 500.  }
\end{figure*}

We first applied GIST (with quantile  $q=0.1$) and  the  robust JAG decomposition.
Figure~\ref{fig:jag_cluster} shows the resulting clusters, where the nodes are placed by the \textit{Fruchterman-Reingold algorithm}~\cite{fruchterman1991graph}. Although no ground truth is available, interestingly, we found that the obtained  10 subnetworks  are highly consistent with the 10 given categories in the data documentation---the corresponding RI is almost as high as  0.9 (cf. Figure~\ref{fig:randIndexSP}).

We then varied $q$ and systematically studied the  clustering results based on GIST. The RIs with respect to the 10 stock categories are shown in Figure~\ref{fig:randIndexSP}.
For comparison, sGTG and sCGD clusterings are also included.  Our JAG decomposition   is quite robust  to the
choice of $q$ in GIST.
It seems that the 10-category structure in the documentation is   reflected on the real stock data. 

\begin{figure*}
  \centering
  \subfloat[\footnotesize Block diagonal screening ]{\label{fig:sth_decomp} 
  \includegraphics[width=0.3\textwidth]{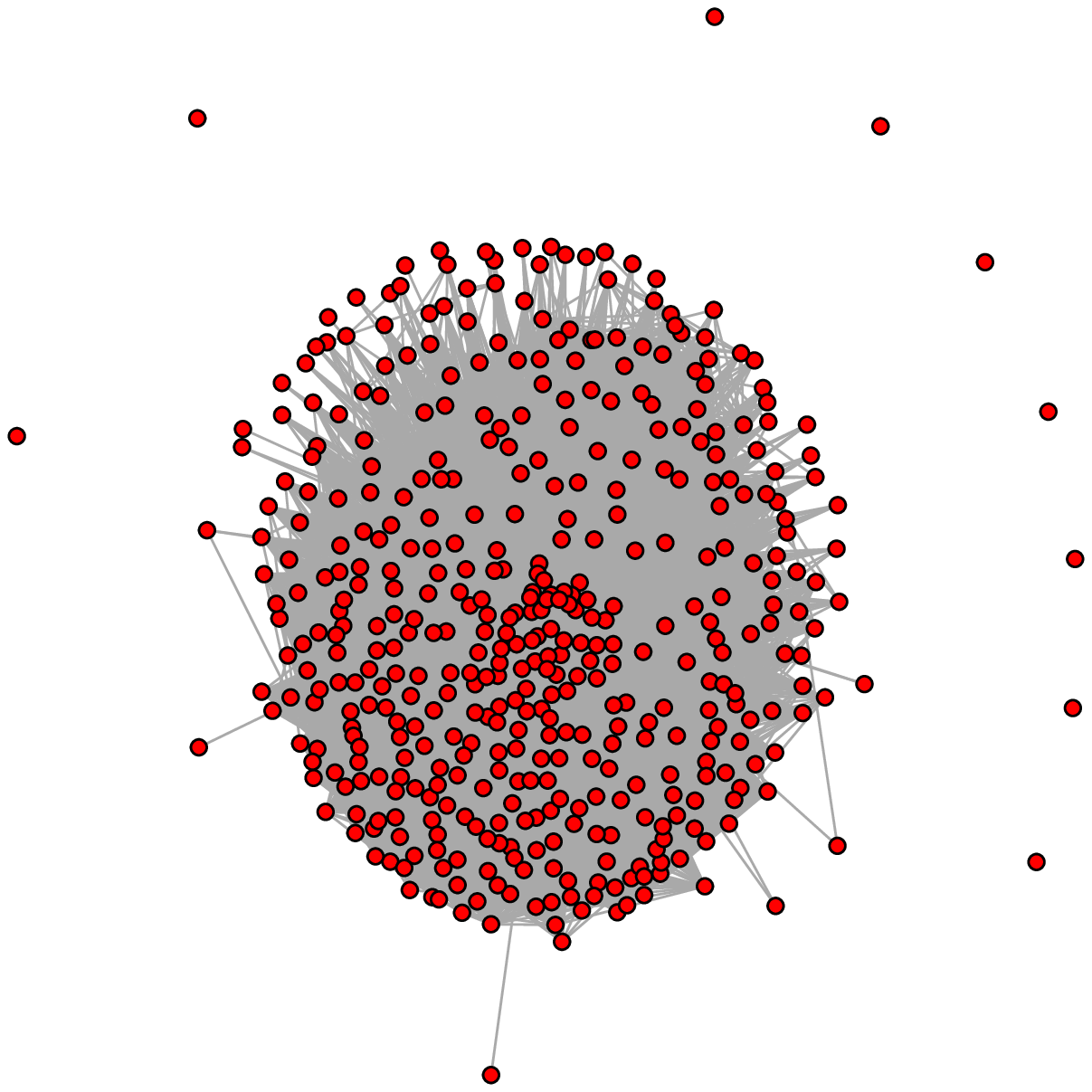}} \qquad  \qquad
  \subfloat[\footnotesize Optimal sCDG estimate ]{\label{fig:huge_opt} 
   \includegraphics[width=0.3\textwidth]{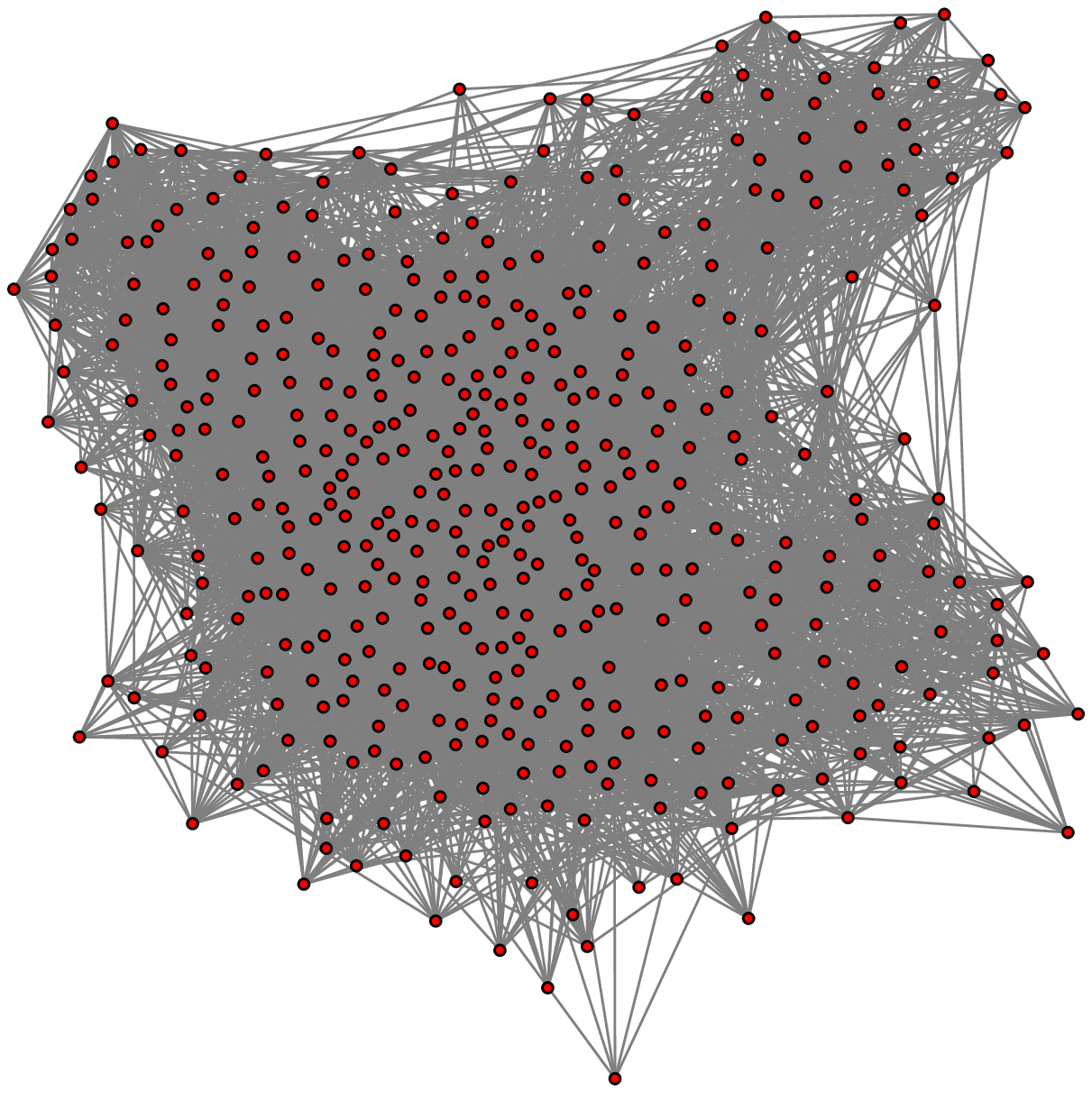}}
  \caption{\footnotesize Gaussian graph learning (or sCDG) on $S\&P$ 500. } \label{fig:sCDGforSP500}
\end{figure*}

We also applied the popular   BDSR \cite{Rahul2012,Witten2011} (which is designed  under the sole CDG learning setup),  to decompose   S\&P 500 into 10 subnetworks. Figure~\ref{fig:sth_decomp} shows that the network is now decomposed into a giant cluster and nine isolated nodes, which is  more difficult to interpret than GIST. Such a decomposition provides little help in reducing the computational cost. Furthermore, Figure~\ref{fig:huge_opt} shows the best tuned  sCDG estimate (using  the R package {\tt huge} \cite{Zhao2012_huge} with default parameters) at    $\lambda^*=0.08$.  To achieve a 10-subnetwork decomposition, we found that $\lambda$ must be greater than or equal to $0.22$.
This is the dilemma discussed  in  Section~\ref{sec:decomposition}:  BDSR resorts to setting an overly large value for $\lambda$ to yield  graph decomposition, while   such a high thresholding level may mask many truly existing edges and result in an inaccurate estimate. Correspondingly,   its  decomposition structure is unreliable.
Of course, the poor performance of BDSR also has a lot to do with the fact that  the transition matrix or GTG estimation is ignored in the sCDG learning. %

We next investigate the forecasting capability of JGSE, by use of the conventional  rolling MSE scheme (see, e.g.,~\cite{Makridakis98_rolling}). Denote the
rolling window size as $W$. Standing at time point $t_0$,
apply the estimation algorithm to the most recent $W$ observations
in the past, i.e., 
$\{\boldsymbol{x}_t\}_{t=t_0-W+1}^{t_0}$. Then use the estimate $\bsb{\hat{B}}_t$  to forecast $\boldsymbol{x}_{t+h}$: 
$\bsb{\hat{x}}_{t+h}
=\bsb{\hat{B}}_t^\tran\bsb{\hat{x}}_{t+h-1}$ for $h\geq 1$, and
$\bsb{\hat{x}}_{t}\eqdef\bsb{x}_{t}$. 
 Repeat the forecasting procedure
 till the rolling window slides to the end of the
time series. The rolling MSE is defined as
$MSE=\frac{1}{n-h-W+1}\sum_{t=W}^{n-h}\|\boldsymbol{x}_{t+h}-\hat{\boldsymbol{x}}_{t+h}\|^2_2$. We set the window size $W=0.8n$
and horizon $h=1$  and compared sGTG, MRCE, and JGSE   in each category. Because of the limited sample size,  the large-data validation used in synthetic experiments is not applicable.   Following \cite{Yuan07,yin2011sparse,guo2011joint}, we chose the tuning parameters by BIC, where the number of degrees of freedom is given by $\sum_{i,j} 1_{\hat{b}_{ij}\neq 0} + \sum_{i\leq j} 1_{\hat{\omega}_{ij}\neq 0} $ if both $\bsbB$ and $\bsbOmega$ are estimated, and $\sum_{i,j} 1_{\hat{b}_{ij}\neq 0}$ if only $\bsbB$ is estimated. Table \ref{tab:rolling_SP500} reports the rolling MSEs  (times 1e+4 for better readability)     for the first five categories. (The conclusions for the last five categories are similar but the
first five have relatively larger dimensions.) Even compared with the widely acknowledged MRCE, JGSE offers better or comparable forecasting performance.

\begin{table*}[htbp]
\begin{center}
\caption{Rolling MSE comparison on S\&P 500.}\label{tab:rolling_SP500}
\begin{tabular}{c|ccccc}
  \hline
  Model \& Method  & Category 1 & Category 2 & Category 3 & Category 4 & Category 5\\ \hline
{sGTG} & 236.6 & 883.5 & 237.4 & 1859.1 & 456.5  \\ \hline
MRCE & 28.4 & 742.6 & 5.0 & 250.1 & 7.9  \\ \hline
{\textbf{JGSE}} & 1.4 & 3.7 & 2.0 & 5.4 & 7.9  \\ \hline
\end{tabular}
\end{center}
\end{table*}

\subsection{NASDAQ-100}
\label{nasdaq}
The NASDAQ-100 consists of 100 of the
largest non-financial companies listed on the NASDAQ stock market.
We collect the closing prices of the stocks for each trading
day from Jan.1 , 2011 to Dec. 31, 2011, which gives  252
samples in total (the data is downloaded from \url{finance.yahoo.com}).
Differencing is applied to remove trends.  There were several significant changes to
the indices during 2011. For example,
NASDAQ
rebalanced the index weights on May 2, 2011 before opening the
market---see \url{http://ir.nasdaqomx.com/releasedetail.cfm?releaseid=561718}. More event details can be found at  \url{http://en.wikipedia.org/wiki/NASDAQ-100#Changes_in_2011}. In consideration of such major changes, we focus on  the following segments.
Segment 1 consists of 62 samples
from Jan. 1 to Apr. 4; Segment 2 consists of 23 samples from Apr. 4
to May 2; Segment 3 consists of 32 samples from May 31 to Jul. 14;
and Segment 4 consists of 98 samples from Jul.15 to Dec. 2.

\begin{figure*}
  \centering
  \subfloat[\footnotesize JAG]{\label{fig:JAG}\includegraphics[width=0.5\textwidth]{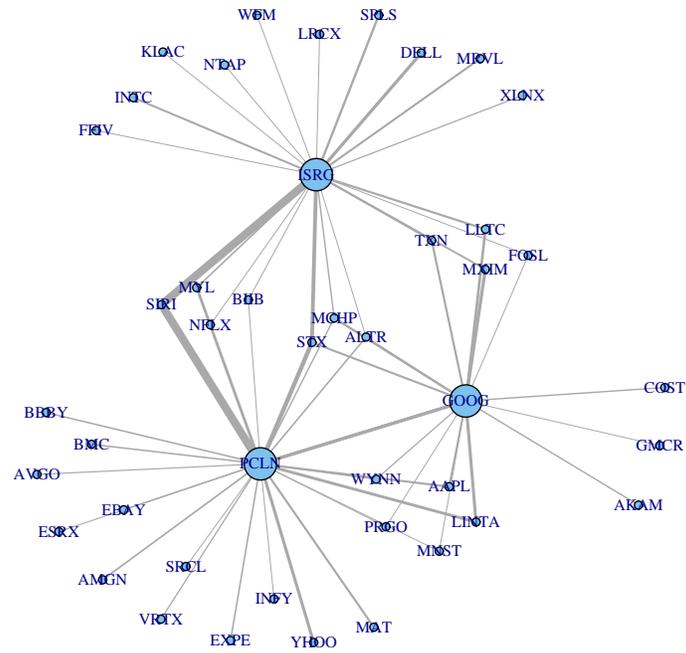}}

  \subfloat[\footnotesize GTG]{\label{fig:causalGraph}\includegraphics[width=0.45\textwidth]{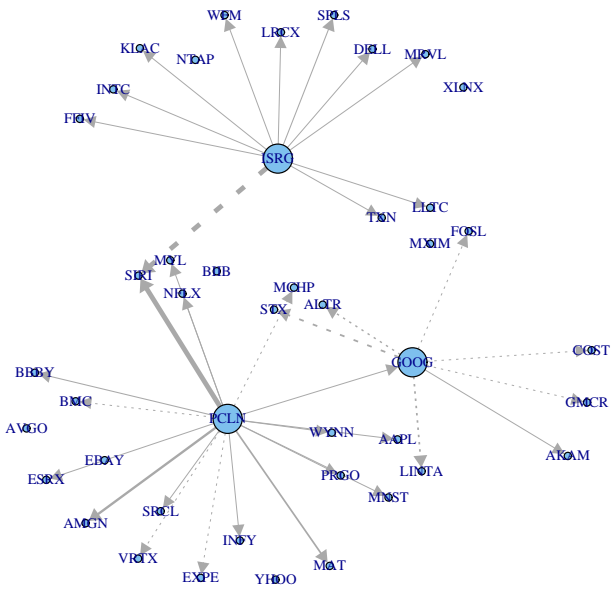}}
  \qquad
  \subfloat[\footnotesize CDG]{\label{fig:dependenceGraph}\includegraphics[width=0.45\textwidth]{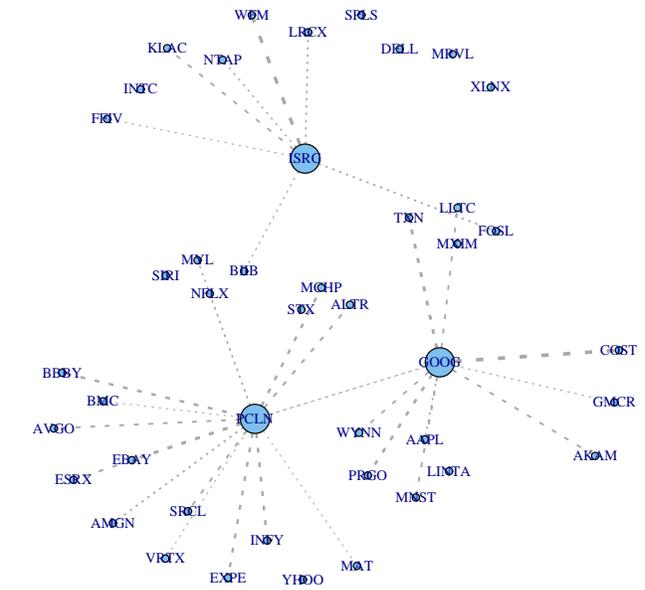}}
  ~ 
  \caption{\footnotesize{The JAG, GTG and CDG of NASDAQ-100. The size of  node is proportional to its degree. The edge width indicates the weight of the connection. Solid/dotted lines represent positive/negative weights.}} \label{fig:causalAndDepenGraph}
\end{figure*}

We  present the analysis of Segment 4 as an example. To get a conservative idea of the network cardinality,  we applied sGTG and sCDG to the data respectively. Sparse graphs are obtained with around $1\%$ connections. We set $q=0.02$ in running the GIST algorithm.
After removing the isolated indices, we applied the FLOG algorithm to obtain  the GTG and CDG estimates. The whole procedure only took a few 
 minutes.  We are particularly interested in the  hub nodes in the JAG. Figure~\ref{fig:causalAndDepenGraph} shows all connections to and from  the hub nodes.  Nicely, the three hubs, PCLN (Priceline.com Inc.), GOOG (Google Inc.) and ISRG (Intuitive Surgical Inc.), come from the three largest sectors of the NASDAQ-100, namely \textit{Consumer Service}, \textit{Technology} and \textit{Health Care}, respectively.
PCLN is a commercial website that helps customers obtain discounts for travel-related
purchases, and it is not surprising  that PCLN is related to  some companies providing similar services, such as EXPE (Expedia Inc.), and some hospitality companies such as WYNN (Wynn Resorts, Limited).  Similarly, GOOG, as a world-famous technology company, is related to many technology based companies, such as AAPL (Apple Inc.), TXN (Texas Instruments Inc.), LLTC (Linear Technology Corporation) and so on. ISRG, a corporation that manufactures robotic surgical systems, is connected with INTC (Intel Corporation),  MRVL (Marvell Technology Group Ltd.) and KLAC (KLA-Tencor Corporation), which  all produce semiconductor chips and nanoelectronic products to be used in  robotics.

The obtained GTG and CDG share some  common connections. For example, PCLN not only has a negative causal influence over EXPE, but shows  negatively correlation  with it conditioned on the other nodes.   On the other hand, the two graphs differ in some ways. For example, although PCLN strongly Granger-causes  SIRI (Sirius XM Holdings, Inc.), they are conditionally independent. The interaction between LLTC and GOOG is  of second order, purely due to their   conditional dependence without any direct Granger causality. Fortunately, JAG encompasses all significant links on either GTG or CDG, and provides comprehensive  network screening.
We have performed similar analysis for other segments and examined the changes of the network topology. Due to page limitation, details are not reported here.

Next, we call the rolling scheme to investigate the forecasting performance  of JGSE.  For comparison,  sGTG was also included;  MRCE is however computationally intractable here, and so we applied our  FLOG$^w$ instead. BIC was used for  regularization parameter tuning.
The rolling MSEs of  three methods are shown in Table~\ref{tab:rolling_NASDAQ},  with window size $W=0.8n$
and horizon $h=1$.   We
see that the joint estimation by FLOG$^w$ outperforms the popular transition estimation (sGTG) in three of the four segments. This suggests the existence of wide range  conditional dependence  between the stocks,  and it is beneficial to take into account such correlations in statistics modeling.  JGSE is able to further improve the forecasting performance by joint regularization, which is however not surprising from Stein et al.'s classical works (e.g., \cite{Stein1956}).
It also has a lot to do with the success of  GIST  in reducing the search space for the fine graph learning. These echo the findings in synthetic data experiments in Section \ref{sec:experiment_accuracy}. 

\begin{table*}
{
  \begin{center}
  \caption{
  Rolling MSE comparison on NASDAQ-100.}\label{tab:rolling_NASDAQ}
\begin{tabular}{l |cccc}
  \hline
  Model \& Method  & Segment 1 & Segment 2 & Segment 3 & Segment 4 \\ \hline
 {sGTG}           & 24.3 & 30.8 & 20.6 & 32.2 \\
FLOG$^w$           & 21.9 & 31.4 & 20.1 & 18.4 \\
{\textbf{{JGSE}}}  & 18.6  & 28.7 & 18.6 & 15.1 \\
  \hline
\end{tabular}
\end{center}
}
\end{table*}

 \section{Conclusion}
 \label{sec:conclusion}
We studied large-scale dynamical networks with sparse first-order and second-order statistical structures, where the first-order connections can be captured by a directed Granger transition graph and the second-order correlations  by an undirected conditional dependence graph. To jointly regularize the two graphs in topology identification and dynamics estimation, we proposed  the 2-stage JGSE framework. The GIST algorithm was developed for JAG screening and decomposition. 
  As demonstrated by  extensive synthetic-data experiments  and real-world applications, our  proposed algorithms beat the commonly used BDSR and MRCE in graph decomposition and estimation.


\bibliographystyle{IEEEtranN}
\bibliography{ref_JAG}

\begin{thebibliography}{47}
\providecommand{\natexlab}[1]{#1}
\providecommand{\url}[1]{#1}
\csname url@samestyle\endcsname
\providecommand{\newblock}{\relax}
\providecommand{\bibinfo}[2]{#2}
\providecommand{\BIBentrySTDinterwordspacing}{\spaceskip=0pt\relax}
\providecommand{\BIBentryALTinterwordstretchfactor}{4}
\providecommand{\BIBentryALTinterwordspacing}{\spaceskip=\fontdimen2\font plus
\BIBentryALTinterwordstretchfactor\fontdimen3\font minus
  \fontdimen4\font\relax}
\providecommand{\BIBforeignlanguage}[2]{{%
\expandafter\ifx\csname l@#1\endcsname\relax
\typeout{** WARNING: IEEEtranN.bst: No hyphenation pattern has been}%
\typeout{** loaded for the language `#1'. Using the pattern for}%
\typeout{** the default language instead.}%
\else
\language=\csname l@#1\endcsname
\fi
#2}}
\providecommand{\BIBdecl}{\relax}
\BIBdecl

\bibitem[Sims(1980)]{Sims1980}
C.~A. Sims, ``Macroeconomics and reality,'' \emph{Econometrica}, vol.~48,
  no.~1, pp. 1--48, Jan. 1980.

\bibitem[Gourieroux and Jasiak(2002)]{gourieroux2002financial}
C.~Gourieroux and J.~Jasiak, ``Financial econometrics problems, models and
  methods,'' \emph{University Presses of California, Columbia and Princeton:
  New Jersey}, 2002.

\bibitem[Fujita et~al.(2007)Fujita, Sato, Garay-Malpartida, Yamaguchi, Miyano,
  Sogayar, and Ferreira]{Fujita07}
A.~Fujita, J.~Sato, H.~Garay-Malpartida, R.~Yamaguchi, S.~Miyano, M.~Sogayar,
  and C.~Ferreira, ``Modeling gene expression regulatory networks with the
  sparse vector autoregressive model,'' \emph{BMC Systems Biology}, vol.~1,
  no.~1, 2007.

\bibitem[Bullmore and Sporns(2009)]{Bullmore09}
E.~Bullmore and O.~Sporns, ``Complex brain networks: graph theoretical analysis
  of structural and functional systems,'' \emph{Nature Reviews Neuroscience},
  vol.~10, pp. 186--198, Mar. 2009.

\bibitem[Granger(1969)]{Granger69}
C.~W.~J. Granger, ``Investigating causal relations by econometric models and
  cross-spectral methods,'' \emph{Econometrica}, vol.~37, no.~3, pp. 424--438,
  1969.

\bibitem[He et~al.(2013)He, She, and Wu]{s2}
Y.~He, Y.~She, and D.~Wu, ``Stationary sparse causality network learning,''
  \emph{J. Mach. Learn. Res.}, vol.~14, pp. 3073--3104, 2013.

\bibitem[She et~al.(2014)She, He, and Wu]{sigmoid}
Y.~She, Y.~He, and D.~Wu, ``Learning topology and dynamics of large recurrent
  neural networks,'' \emph{IEEE Transactions on Signal Processing}, 2014, to
  appear.

\bibitem[Stein(1956)]{Stein1956}
C.~Stein, ``Inadmissibility of the usual estimator for the mean of a
  multivariate distribution,'' \emph{Proc. Third Berkeley Symp. Math. Statist.
  Prob.}, vol.~1, pp. 197--206, 1956.

\bibitem[Bolstad et~al.(2011)Bolstad, Van~Veen, and Nowak]{bolstad2011causal}
A.~Bolstad, B.~D. Van~Veen, and R.~Nowak, ``Causal network inference via group
  sparse regularization,'' \emph{Signal Processing, IEEE Transactions on},
  vol.~59, no.~6, pp. 2628--2641, 2011.

\bibitem[Valdes-Sosa(2005)]{Pedro05}
P.~A. Valdes-Sosa, ``Estimating brain functional connectivity with sparse
  multivariate autoregression,'' \emph{Phil. Trans. R. Soc. B}, pp. 969--981,
  2005.

\bibitem[Banerjee et~al.(2008)Banerjee, El~Ghaoui, and
  d'Aspremont]{banerjee2008model}
O.~Banerjee, L.~El~Ghaoui, and A.~d'Aspremont, ``Model selection through sparse
  maximum likelihood estimation for multivariate gaussian or binary data,''
  \emph{The Journal of Machine Learning Research}, vol.~9, pp. 485--516, 2008.

\bibitem[Friedman et~al.(2008)Friedman, Hastie, and
  Tibshirani]{Friedman08_graphLasso}
J.~Friedman, T.~Hastie, and R.~Tibshirani, ``Sparse inverse covariance
  estimation with the graphical lasso,'' \emph{Biostatistics}, vol.~9, no.~3,
  pp. 432--441, Jul. 2008.

\bibitem[Bickel and Levina(2008)]{Bickel2008}
J.~Bickel and E.~Levina, ``Regularized estimation of large covariance
  matrices,'' \emph{Ann. Statist}, vol.~36, no.~1, pp. 199--227, 2008.

\bibitem[Meinshausen and B{\"u}hlmann(2006)]{meinshausen2006high}
N.~Meinshausen and P.~B{\"u}hlmann, ``High-dimensional graphs and variable
  selection with the lasso,'' \emph{The Annals of Statistics}, vol.~34, no.~3,
  pp. 1436--1462, 2006.

\bibitem[Rothman et~al.(2010)Rothman, Levina, and
  Zhu]{Rothman2010_SparseBsparseTheta}
A.~J. Rothman, E.~Levina, and J.~Zhu, ``Sparse multivariate regression with
  covariance estimation,'' \emph{Journal of Computational and Graphical
  Statistics}, vol.~19, no.~4, 2010.

\bibitem[Lee and Liu(2012)]{lee2012simultaneous}
W.~Lee and Y.~Liu, ``Simultaneous multiple response regression and inverse
  covariance matrix estimation via penalized gaussian maximum likelihood,''
  \emph{Journal of Multivariate Analysis}, 2012.

\bibitem[James and Stein(1961)]{james1961}
W.~James and C.~Stein, ``Estimation with quadratic loss,'' Berkeley, Calif.,
  pp. 361--379, 1961.

\bibitem[Fink et~al.(2009)Fink, Grabner, Benedek, Reishofer, Hauswirth, Fally,
  Neuper, Ebner, and Neubauer]{Fink2009}
A.~Fink, R.~H. Grabner, M.~Benedek, G.~Reishofer, V.~Hauswirth, M.~Fally,
  C.~Neuper, F.~Ebner, and A.~C. Neubauer, ``The creative brain: Investigation
  of brain activity during creative problem solving by means of eeg and fmri,''
  \emph{Human Brain Mapping}, vol.~30, no.~3, pp. 734--748, 2009.

\bibitem[Stock and Watson(2012)]{stock2012generalized}
J.~H. Stock and M.~W. Watson, ``Generalized shrinkage methods for forecasting
  using many predictors,'' \emph{Journal of Business \& Economic Statistics},
  vol.~30, no.~4, pp. 481--493, 2012.

\bibitem[Witten et~al.(2011)Witten, Friedman, and Simon]{Witten2011}
D.~M. Witten, J.~H. Friedman, and N.~Simon, ``New insights and faster
  computations for the graphical lasso,'' \emph{Journal of Computational and
  Graphical Statistics}, vol.~20, no.~4, pp. 892--900, 2011.

\bibitem[Mazumder and Hastie(2012)]{Rahul2012}
R.~Mazumder and T.~Hastie, ``Exact covariance thresholding into connected
  components for large-scale graphical lasso,'' \emph{J. Mach. Learn. Res.},
  vol.~13, pp. 781--794, Mar. 2012.

\bibitem[Zhao et~al.(2012)Zhao, Liu, Roeder, Lafferty, and
  Wasserman]{Zhao2012_huge}
T.~Zhao, H.~Liu, K.~Roeder, J.~Lafferty, and L.~Wasserman, ``The huge package
  for high-dimensional undirected graph estimation in r,'' \emph{J. Mach.
  Learn. Res.}, vol.~13, no.~12, pp. 1059--1062, Jun. 2012.

\bibitem[L\"{u}tkepohl(2007)]{LutkBook}
H.~L\"{u}tkepohl, \emph{New Introduction to Multiple Time Series Analysis},
  1st~ed.\hskip 1em plus 0.5em minus 0.4em\relax Springer, Oct. 2007.

\bibitem[Tibshirani(1996)]{Tibshirani96}
R.~Tibshirani, ``Regression shrinkage and selection via the lasso,''
  \emph{Journal of the Royal Statistical Society}, vol.~58, no.~1, pp.
  267--288, 1996.

\bibitem[Yuan and Lin(2007)]{Yuan07}
M.~Yuan and Y.~Lin, ``Model selection and estimation in the gaussian graphical
  model,'' \emph{Biometrika}, vol.~94, no.~1, pp. 19--35, 2007.

\bibitem[Efron and Morris(1973)]{Efron1973}
B.~Efron and C.~Morris, ``\BIBforeignlanguage{English}{Stein's estimation rule
  and its competitors--an empirical bayes approach},''
  \emph{\BIBforeignlanguage{English}{Journal of the American Statistical
  Association}}, vol.~68, no. 341, pp. pp. 117--130, 1973.

\bibitem[She(2009)]{She09}
Y.~She, ``Thresholding-based iterative selection procedures for model selection
  and shrinkage,'' \emph{Electron. J. Statist}, vol.~3, pp. 384--415, 2009.

\bibitem[She(2012)]{SheGLMTISP}
------, ``An iterative algorithm for fitting nonconvex penalized generalized
  linear models with grouped predictors,'' \emph{Computational Statistics and
  Data Analysis}, vol.~9, pp. 2976--2990, 2012.

\bibitem[Fan and Li(2001)]{Fan01_SCAD}
J.~Fan and R.~Li, ``Variable selection via nonconcave penalized likelihood and
  its oracle properties,'' \emph{Journal of the American Statistical
  Association}, vol.~96, pp. 1348--1360, Dec. 2001.

\bibitem[She et~al.(2013)She, Li, Wang, and Wu]{SheSpec}
Y.~She, H.~Li, J.~Wang, and D.~Wu, ``Grouped iterative spectrum thresholding
  for super-resolution sparse spectrum selection,'' \emph{IEEE Transactions on
  Signal Processing}, vol.~61, pp. 6371--6386, 2013.

\bibitem[She(2013)]{SheTISPMat}
Y.~She, ``Reduced rank vector generalized linear models for feature
  extraction,'' \emph{Statistics and Its Interface}, vol.~6, pp. 197--209,
  2013.

\bibitem[She()]{sel-rrr}
------, ``Selectable factor extraction in high dimensions,'' arXiv:1403.6212.

\bibitem[Armijo(1966)]{Armi66a}
L.~Armijo, ``Minimization of functions having {L}ipschitz continuous first
  partial derivatives,'' \emph{Pacific Journal of Mathematics}, vol.~16, no.~1,
  pp. 1--3, 1966.

\bibitem[Duchi et~al.(2008)Duchi, Gould, and Koller]{Duchi08_BlockSparsInvCov}
J.~Duchi, S.~Gould, and D.~Koller, ``Projected subgradient methods for learning
  sparse {G}aussians,'' in \emph{Proceedings of the Twenty-fourth Conference on
  Uncertainty in AI (UAI)}, 2008.

\bibitem[Dulmage and Mendelsohn(1958)]{dulmage1958coverings}
A.~L. Dulmage and N.~S. Mendelsohn, ``Coverings of bipartite graphs,''
  \emph{Canadian Journal of Mathematics}, vol.~10, no.~4, pp. 516--534, 1958.

\bibitem[von Luxburg(2007)]{Von2006tutorial_spectral}
U.~von Luxburg, ``A tutorial on spectral clustering,'' \emph{Statistics and
  Computing}, vol.~17, no.~4, pp. 395--416, 2007.

\bibitem[von Luxburg et~al.(2008)von Luxburg, Belkin, and
  Bousquet]{von2008consistency}
U.~von Luxburg, M.~Belkin, and O.~Bousquet, ``Consistency of spectral
  clustering,'' \emph{The Annals of Statistics}, pp. 555--586, 2008.

\bibitem[Fraley and Raftery(1998)]{fraley1998many}
C.~Fraley and A.~E. Raftery, ``How many clusters? which clustering method?
  answers via model-based cluster analysis,'' \emph{The Computer Journal},
  vol.~41, no.~8, pp. 578--588, 1998.

\bibitem[Tibshirani et~al.(2001)Tibshirani, Walther, and
  Hastie]{tibshirani2001estimating}
R.~Tibshirani, G.~Walther, and T.~Hastie, ``Estimating the number of clusters
  in a data set via the gap statistic,'' \emph{Journal of the Royal Statistical
  Society: Series B (Statistical Methodology)}, vol.~63, no.~2, pp. 411--423,
  2001.

\bibitem[Sugar and James(2003)]{sugar2003finding}
C.~A. Sugar and G.~M. James, ``Finding the number of clusters in a dataset: An
  information-theoretic approach,'' \emph{Journal of the American Statistical
  Association}, vol.~98, no. 463, pp. 750--763, 2003.

\bibitem[Nocedal and Wright(2006)]{nocedal2006numerical}
J.~Nocedal and S.~J. Wright, \emph{Numerical Optimization}, 2nd~ed.\hskip 1em
  plus 0.5em minus 0.4em\relax New York: Springer, 2006.

\bibitem[Friedman et~al.(2007)Friedman, Hastie, H{\"o}fling, and
  Tibshirani]{friedman2007pathwise}
J.~Friedman, T.~Hastie, H.~H{\"o}fling, and R.~Tibshirani, ``Pathwise
  coordinate optimization,'' \emph{The Annals of Applied Statistics}, vol.~1,
  no.~2, pp. 302--332, 2007.

\bibitem[Rand(1971)]{Rand_randIndex}
W.~M. Rand, ``Objective criteria for the evaluation of clustering methods,''
  \emph{Journal of the American Statistical Association}, vol.~66, no. 336, pp.
  846--850, 1971.

\bibitem[Fruchterman and Reingold(1991)]{fruchterman1991graph}
T.~M. Fruchterman and E.~M. Reingold, ``Graph drawing by force-directed
  placement,'' \emph{Software: Practice and experience}, vol.~21, no.~11, pp.
  1129--1164, 1991.

\bibitem[Makridakis et~al.(1998)Makridakis, Wheelwright, and
  Hyndman]{Makridakis98_rolling}
Makridakis, Wheelwright, and Hyndman, \emph{Forecasting: Methods and
  Applications}.\hskip 1em plus 0.5em minus 0.4em\relax Wiley, 1998.

\bibitem[Yin and Li(2011)]{yin2011sparse}
J.~Yin and H.~Li, ``A sparse conditional gaussian graphical model for analysis
  of genetical genomics data,'' \emph{The Annals of Applied Statistics},
  vol.~5, no.~4, p. 2630, 2011.

\bibitem[Guo et~al.(2011)Guo, Levina, Michailidis, and Zhu]{guo2011joint}
J.~Guo, E.~Levina, G.~Michailidis, and J.~Zhu, ``Joint estimation of multiple
  graphical models,'' \emph{Biometrika}, p. asq060, 2011.

\end{thebibliography}

\end{document}